\documentclass[useAMS,usenatbib,usegraphicx]{mn2e}

\title[Periodic changes in ring morphology]
{Periodic changes in the morphology of the Galactic resonance rings}
\author[Melnik et al.]{A.M.~Melnik$^1$\thanks{E-mail:
anna@sai.msu.ru}, E.N.~Podzolkova$^{1, 2}$, A.K.~Dambis$^1$,\\
$^1$Sternberg Astronomical Institute, Lomonosov Moscow State
University, Universitetskii pr. 13, Moscow, 119991, Russia \\
$^2$Faculty of Physics, Lomonosov Moscow State University, Leninskie
Gory 1-2, Moscow, 119991, Russia}

\begin{document}

\date{Accepted 2023 December 00. Received 2022 December 00; in original form 2023 December 00}


\maketitle

\label{firstpage}

\begin {abstract}
We study the periodic  enhancement  of either trailing or leading
segments of the resonance elliptical rings in the dynamical model of
the Galaxy which reproduces distributions of observed velocities
derived from {\it Gaia} DR3 (EDR3) data  along the Galactocentric
distance. The model disc forms a nuclear ring, an inner combined ring
and outer resonance rings $R_1$ and $R_2$. The backbone of the inner
combined ring is banana-type orbits around the Lagrange equilibrium
points $L_4$ and $L_5$. Orbits associated with the unstable
equilibrium points $L_1$ and $L_2$ also support the inner ring. We
have found the changes of the morphology of the inner ring with a
period of $P=0.57\pm0.02$ Gyr, which  is close to the period of
revolution along the long-period orbits around the points $L_4$ and
$L_5$.  A possible explanation of these morphological changes is the
formation of an overdensity which  then begins circulating along the
closed contour. In the region of the Outer Lindblad Resonance (OLR),
we have found the changes of the morphology of the outer rings with a
period of $P=2.0\pm0.1$ Gyr. Probably, the morphological changes of
the outer rings are due to the orbits trapped by the OLR. These
orbits exhibit librations of the direction of orbital elongation with
respect to the minor axis of the bar as well as the long-term
variations in the stellar angular momentum, energy, average radius of
the orbit, and eccentricity. Among many librating orbits, we
discovered orbits with the libration period of $P=1.91\pm0.01$ Gyr,
which may cause the morphological changes of the outer rings.
\end{abstract}


\maketitle

\section{Introduction}

There is extensive  evidence for the presence of the bar in our
Galaxy \citep{dwek1995,   fux2001, muhlbauer2003, benjamin2005,
cabrera-lavers2007, churchwell2009, gerhard2011, gonzalez2012,
nesslang2016, simion2017}. Solid rotation of the bar in the
differentially rotating disc causes the appearance of resonances and
resonance elliptical rings \citep{buta2017}.

The positions of the resonances in a galactic disc are determined
from the condition:

\begin{equation}
\frac{m}{n}= \frac{\kappa}{\Omega-\Omega_b}, \label{resonance}
\end{equation}

\noindent where $m$ is the number of full epicyclic oscillations that
a star makes during $n$ revolutions relative to the bar. Usually the
case $n=1$ is considered. The ratio $m/n=+2/1$ determines the
position of the Inner Lindblad Resonance (ILR), and the ratio
$m/n=-2/1$ determines the radius of the Outer Lindblad resonance
(OLR). The resonances $m/n=\pm$4/1 are also of great interest
\citep{contopoulos1983, contopoulos1989, athanassoula1992a}.

Resonance elliptical rings in barred galaxies arise due to adjustment
of  epicyclic motions of  stars in accordance with their orbital
rotation with respect to the bar. There are three basic types of
rings:  nuclear ($n$), inner ($r$), and outer ($R_1$, and $R_2$)
rings. Nuclear rings are  usually located near the ILR of the bar and
are elongated perpendicular to the bar,  inner rings lie near the
Corotation Radius (CR) and are usually elongated parallel to the bar,
outer rings are located near the OLR and have two possible
orientations: the rings $R_1$ and pseudo-rings (broken rings) $R_1'$
are aligned perpendicular to the bar, but the rings $R_2$ and
pseudo-rings $R_2'$ are elongated parallel to the bar. From the two
types of outer rings,  ring $R_1$ is located a bit closer to the
Galactic center than  ring $R_2$ \citep{schwarz1981, buta1995,
buta1996, buta1991, byrd1994, rautiainen1999, rautiainen2000,
rodriguez2008,sormani2018}. \citet{rautiainen2000} also found cyclic
or semi-cyclic variations in the morphology of the outer rings: from
$R_2$ to $R_1R_2$ and back. The fraction of systems with outer rings
among galaxies with a moderate or a strong bar is as high as 20--30
per cent \citep{comeron2014}.

In addition to the resonance elliptical rings, there are also
lenticular structures that are very similar to the rings but have
flatter density distributions, especially at the outer boundary.
These can have the form of inner ($l$) and outer lenses ($L$) whose
orientations in most cases coincide with the orientations of the
inner ($r$) and outer ($R_1$) rings, respectively
\citep{kormendy1979, athanassoula1982, buta1996, laurikainen2011}.

The backbone of resonance rings are direct stable periodic orbits
which are followed by  a large number of stars in quasi-periodic
orbits. Inside the CR  family of periodic orbits $x_1$ is elongated
along the bar and form its backbone. Near the OLR, the main family of
periodic orbits $x_1$ splits into two families: $x_1(1)$ and
$x_1(2)$. The periodic orbits  $x_1(2)$ are elongated perpendicular
to the bar and lie between the resonances $-4/1$ and OLR while the
periodic orbits $x_1(1)$ are elongated parallel to the bar and are
located outside the OLR.  In the case of a weak bar, the stable
periodic orbits $x_1(2)$ and $x_1(1)$ support the rings $R_1$ and
$R_2$, respectively \citep{contopoulos1980, contopoulos1989,
schwarz1981, buta1996}.

In the case of a strong bar, the periodic orbits $x_1(2)$ lying
between the resonances $-4/1$ and OLR become unstable
\citep{contopoulos1980, contopoulos1989}. However, even in this case,
spiral arms and ring-like structures can be supported for some time
by sticky-chaotic orbits  \citep{contopoulos_harsoula2010} associated
with the unstable equilibrium points $L_1$ and $L_2$ located at the
ends of the bar \citep{athanassoula2009a, athanassoula2009b,
athanassoula2010}.

In barred galaxies, there are Lagrange equilibrium points at which
the bar gravity is balanced by the centrifugal inertial force. The
Lagrange equilibrium points $L_1$, $L_2$, $L_4$ and $L_5$ are located
near the CR on the minor ($L_4$ and $L_5$) and major ($L_1$ and
$L_2$)  axes of the bar. In the neighborhood of the stable
equilibrium points $L_4$ and $L_5$, there are two types of periodic
orbits:  short-period  (SPO) and long-period  (LPO) orbits.
Short-period orbits are similar to a small ellipse and their orbital
period is close to the epicyclic one while  long-period orbits
resemble a banana and their orbital period is several times longer.
Most non-periodic banana-type orbits are a combination of short- and
long-period oscillations  \citep{contopoulos1978, contopoulos1983,
contopoulos1989, binney2008}.

A concentration of stars near the equilibrium points L4 and L5 can
also support spiral arms \citep{barbanis1970} and ring-like
structures, provided the stars have the appropriate energy and are
not trapped around them \citep{danby1965, patsis2017}.

The periodic  enhancement  of either trailing or leading segments of
the resonance elliptical structures can be found in many studies. We
can clearly see the periodic motion of the density maxima along
spiral arms emanating from the bar ends in the study by \citet[][Fig.
14 therein]{contopoulos2009}.  Furthermore, the stellar response of
the model with the Ferrers bar in \citet[][]{athanassoula1992b}
demonstrates  the predominance of leading or trailing spiral arms at
different time instants.  On the other hand, the gas response always
shows the formation of the overdensity on the leading side of the
bar.

The notion of resonance as a certain radius is true only for circular
orbits, but for elliptical orbits, we deal with a Lindblad Zone
\citep{struck2015a, struck2015b}.

Stellar motions near the bar resonances  are often described in terms
of angle-action variables \citep{weinberg1994}. It is often
convenient to consider the frequency of the resonance:

\begin{equation}
\Omega_s=n\Omega_R+m(\Omega-\Omega_b),
 \label{omega_s}
\end{equation}

\noindent where $\Omega_R$ is the frequency of radial oscillations.
In the epicyclic approximation $\Omega_R$ coincides with the
epicyclic frequency, $\Omega_R=\kappa$. The value $n=0$ corresponds
to the Corotation Resonance, but the combination of $n=\pm1$ and
$m=2$ describes the Lindblad resonances. The integral of $\Omega_s$
over the time determines the so-called slow angle variable

\begin{equation}
\theta_s=\int_{0}^{t}\Omega_s(t')dt',
 \label{theta_s}
\end{equation}

\noindent which varies very slowly near the resonances. The slow
angle corresponds to the precession angle of the orbit relative to
the bar, whereas the fast angle variable, $\theta_f$, describes the
motion of the star around its orbit. \citet{weinberg1994} showed that
the slow angle $\theta_s$ can change in two ways: the direction of
orbit elongation may oscillate within certain angles or may rotate
without any restriction on angles. The study of orbits near the
resonances of the bar became especially salient in the last decade
\citep{struck2015b, monari2017, trick2021, chiba2021,
chiba_schonrich2021, chiba_schonrich2022}.

The studies carried out by our team  provided extensive evidence for
the presence of the Galactic outer resonance rings located near the
solar circle. We developed models of the Galaxy with the outer
resonance rings $R_1$ and $R_2$ which reproduce well the kinematics
of  young objects (OB-associations, young clusters, classical
Cepheids) in the 3-kpc solar neighborhood
\citep{melnikrautiainen2009, rautiainen2010, melnik2015, melnik2016}.
The velocity dispersion of these objects in the Galactic plane is
$\sim 10$ km s$^{-1}$. Our models can also explain the distribution
of  star-forming regions in the Galactic disc \citep{melnik2011}. The
best agreement between model and observed velocities of OB
associations corresponds to the model with the bar angular velocity
of $\Omega_b=50\pm2$ km s$^{-1}$ kpc$^{-1}$ \citep{melnik2019}.

In our latest study \citep{melnik2021}, we compared the model and
observed velocity distributions of   old disc stars along the
Galactocentric distance $R$.  The radial velocity dispersion of these
stars at the solar Galactocentric distance is $\sim30$ km s$^{-1}$.
The best agreement between the model and observed velocities is
achieved for the model with the bar angular velocity of
$\Omega_b=55\pm3$ km s$^{-1}$ kpc$^{-1}$, the position angle of the
bar with respect to the Sun of $\theta_b=45\pm15^\circ$ and the age
of the Galactic bar of $1.8\pm0.5$ Gyr.

Any elliptical ring can be divided into 4 similar segments each of
which can be represented as a fragment of a trailing ($R$ decreases
with increasing  $\theta$) or a leading ($R$ increases with
increasing $\theta$) spiral, where the galactocentric angle $\theta$
increases in the sense of the galactic rotation.

In this paper, we study periodic changes in the morphology of
resonance rings using the Galactic model from \citet{melnik2021}. We
have found the periodic enhancement of either trailing or leading
segments in the region of the inner combined ring and in the region
of the outer rings which have different periods and are likely to
appear due to different causes.

In Section 2 we compare the distributions of observed velocities
along the distance $R$ derived from  the {\it Gaia} DR3 and {\it
Gaia} EDR3 data. Section 3 describes the dynamical model of the
Galaxy. In section 4 we analyze periodic  changes in the stellar
density in the region of the inner ring and the outer rings. In
Section 5 we study orbits that can support the inner combined ring
and propose a mechanism producing the periodic changes in the
morphology of the inner ring. In Section 6 we study the orbits near
the OLR of the bar and show that the librations of the direction of
orbit elongation can produce changes in the morphology of the outer
rings. The discussion and main conclusions are given in Section 7.

\begin{figure*}
\resizebox{\hsize}{!}{\includegraphics{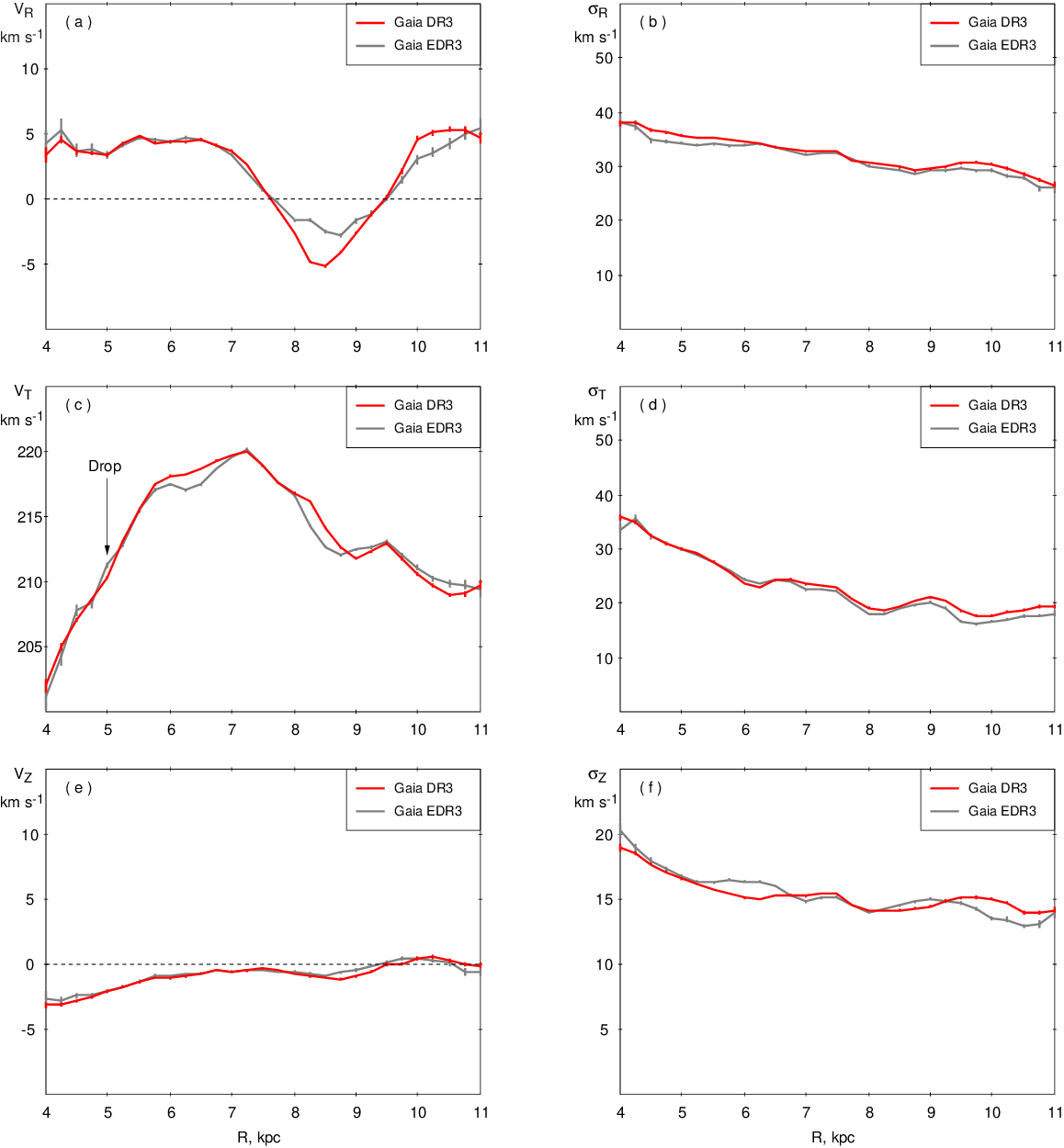}} \caption{(Left
panel) Dependences of the radial $V_R$ (a), azimuthal $V_T$ (c), and
vertical $V_Z$ (e) components of the median velocity on
Galactocentric distance $R$ derived from  {\it Gaia} DR3 (red lines)
and {\it Gaia} EDR3 data (gray lines) in $\Delta R=250$-pc wide bins.
The vertical lines on the curves indicate random errors in the
determination of the median velocities. The arrow on  frame (c)
indicates the drop in the median velocity $V_T$ which seems to be due
to  selection effects. (Right panel)  Dependences of the median
dispersion of the radial $\sigma_R$ (b), azimuthal $\sigma_T$ (d) and
vertical $\sigma_z$ (f) velocities on Galactocentric distance $R$.
The distributions of the velocities $V_R$, $V_T$, $V_Z$ and the
velocity dispersions $\sigma_R$, $\sigma_T$, and $\sigma_Z$ derived
from {\it Gaia} DR3 and  {\it Gaia} EDR3 data can be seen to
generally agree well with each other.} \label{obs_prof}
\end{figure*}
\begin{figure*}
\resizebox{\hsize}{!}{\includegraphics{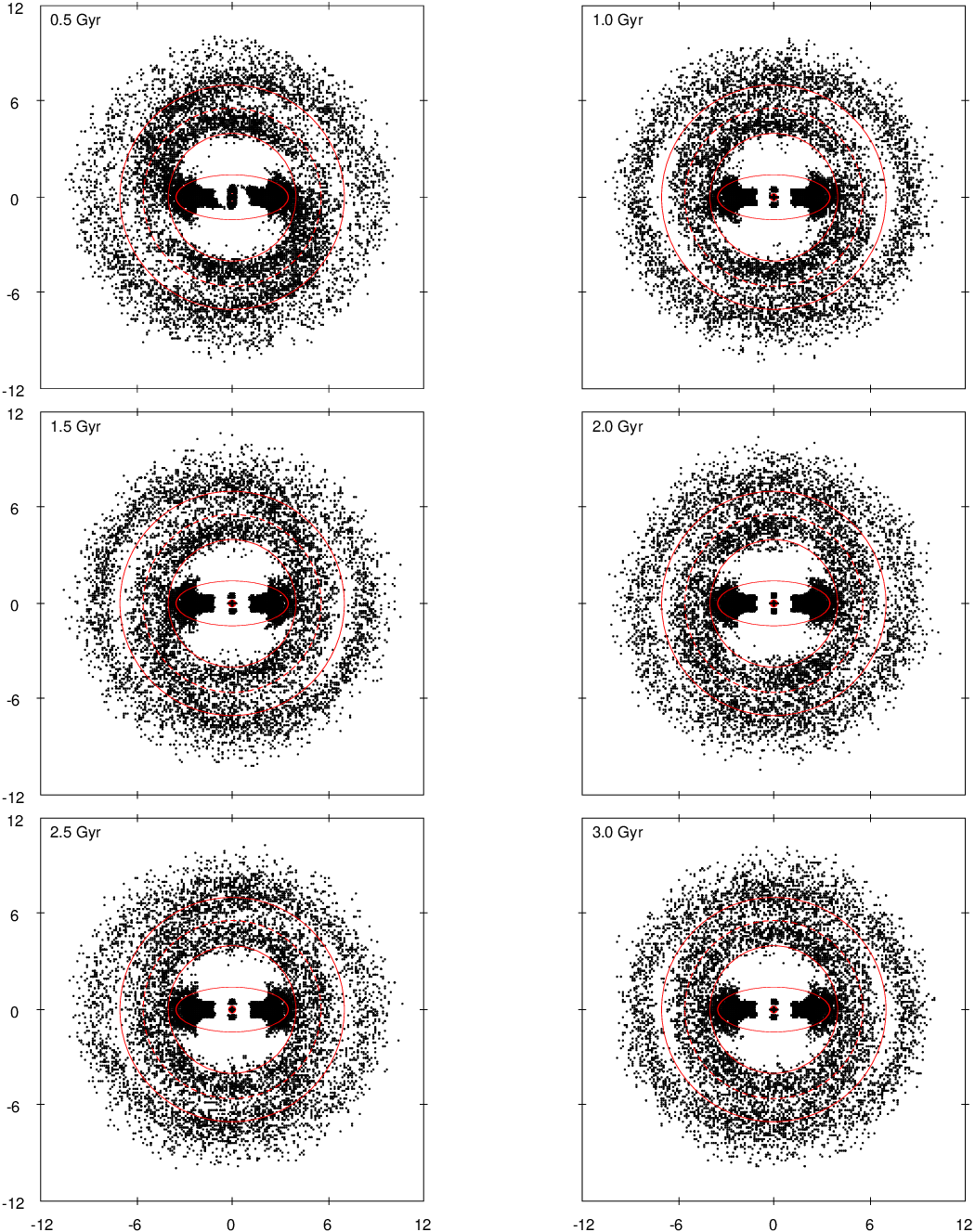}}
\caption{Distributions of stars in the Galactic plane at the time
instants 0.5, 1.0, 1.5, 2.0, 2.5, and 3.0 Gyr processed by the
program that increases contrast. The parameter $h$, which controls
the contrast, is $h=1.5$. In these plots 10 per cent of stars  are
shown. The ellipse shows the position of the bar. The red solid lines
indicate the positions of the CR and OLR, and the dashed line shows
the position of the  resonance $-4/1$. The Galaxy rotates
counterclockwise. The  size of each frame is $24\times24$ kpc$^2$.}
\label{distrib}
\end{figure*}

\section{Observations}

In our latest study \citep{melnik2021}, we selected stars from the
{\it Gaia} EDR3 catalogue \citep{prusti2016, brown2021,
lindegren2021} that lie near the plane of the Galaxy $|z|<200$ pc and
in the sector of the Galactocentric angles
$|\theta-\theta_0|<15^\circ$, where $\theta_0$ is the azimuthal angle
of the Sun with respect to the bar major axis ($\theta_0=-\theta_b$),
and calculated the median radial, $V_R$, azimuthal, $V_T$, and
vertical, $V_Z$, velocity components in $\Delta R=250$-pc wide bins
along the distance $R$. The final sample included $2.39\times 10^6$
{\it Gaia} EDR3 stars. Our model appeared to reproduce well the
observed dependences of the velocities $V_R$ and $V_T$ on $R$ in the
distance range $|R-R_0|<1.5$ kpc, where $R_0$ is the Galactocentric
distance of the Sun.

We adopt the solar Galactocentric distance  $R_0=7.5$ kpc
\citep[][]{glushkova1998, nikiforov2004, eisenhauer2005,
nishiyama2006, feast2008, groenewegen2008, reid2009b, dambis2013,
francis2014, boehle2016, branham2017, iwanek2023}. Generally, the
choice of the value of $R_0$ within the limits 7--9 kpc has
practically no effect on the results.

In this study we calculate the dependences of $V_R$, $V_T$, and $V_Z$
on $R$ for {\it Gaia} DR3 data \citep[][]{vallenari2022}. The final
sample includes $8.32\times 10^6$ {\it Gaia} DR3 stars.
Fig.~\ref{obs_prof} shows the distributions of the velocities $V_R$,
$V_T$, and $V_Z$ and those of the velocity dispersions $\sigma_R$,
$\sigma_T$, and $\sigma_Z$ derived from {\it Gaia} DR3 and {\it Gaia}
EDR3 data. Average differences between the velocities derived from
{\it Gaia} EDR3 and {\it Gaia} DR3 data do not exceed 1 km s$^{-1}$.
The main difference is observed in the $V_R$-velocity profile
(Fig.~\ref{obs_prof}a), which has a minimum at $R=8.5$ kpc with a
depth of $-5.2$ and $-2.8$ km s$^{-1}$ according to {\it Gaia} DR3
and {\it Gaia} EDR3 data, respectively. However, this difference does
not change the general tendency: the smooth fall  of radial
velocities near the solar circle.

The radial profile of the velocity dispersion $\sigma_R$ derived from
{\it Gaia} DR3 data in the distance interval $R=5$--11 kpc can be
approximated by the exponential law with the scale length of
$S_R=23.7\pm1.6$ kpc, which agrees with the estimate $S_R=22.3\pm1.4$
kpc derived from {\it Gaia} EDR3 data.

Note  an important feature, which has received  little attention,
namely -- a sharp drop in the azimuthal velocity $V_T$ at $R\approx
4.0$--5.5 kpc (Fig.~\ref{obs_prof}c). This effect is likely due to
the lack of stars associated with the thin disc compared to the
fraction of stars associated with the thick disc and halo in this
region. Extinction in  Galactic midplane increases very rapidly
towards the Galactic center \citep{neckel1980, marshall2006,
melnik2015, melnik2016}. The sources with line-of-sight velocity
listed in the {\it Gaia} DR3 ({\it Gaia} EDR3) catalogue are mostly
brighter than $G=14^m$ ($G=13^m$) \citep{vallenari2022,brown2018,
sartoretti2018,katz2018,katz2019}. As extinction increases,  stars
associated with the thick disc and halo are more likely to have
line-of-sight velocity measurements compared to thin-disc  stars.
There are two reasons for this: the distribution along $z$ and
colour. Thin-disc stars strongly concentrate to the Galactic
midplane, where extinction is stronger compared to thick-disc and
halo stars, and therefore  the fraction of thick-disc and halo stars
in the $|z|<200$-pc layer that are brighter than $G=14^m$ must
increase with increasing extinction. In addition, extinction has
stronger effect on blue stars than on red ones, and therefore the
fraction  of red stars brighter than $G=14^m$ must increase with
increasing extinction. Thus, the sharp drop in the velocity $V_T$
towards the Galactic center seems to be caused by selection effects.

Our models reproduce well the drop in the velocity $V_T$ at
Galactocentric distances $R=7.5$--10 kpc. This region lies in the
direction towards the Galactic anticenter, where extinction is small.
Here the decrease in the azimuthal velocity $V_T$ is caused by the
influence of the bar.

Note that the median velocities $V_Z$ in the direction perpendicular
to the Galactic plane derived from the {\it Gaia} DR3 data do not
exceed $|V_Z|<3$ km s$^{-1}$ in the distance interval $R=4$--11 kpc
and do not exceed $|V_Z|<1$ km s$^{-1}$ in the distance interval
$|R-R_0|<1.5$. Quite strong variations of both $V_R$ and $V_T$ over
the Galactocentric distance interval $R=6$--10 kpc combined with the
absence of noticeable systematic motions in $Z$-direction suggest
that stellar motions in the Galactic plane and in the vertical
direction are independent. This justifies our use of a 2D-model of
the Galaxy.

\section{Model}

We used a 2D-model of the Galaxy with analytical potential and
studied the response of the stellar disc to the bar perturbation. Our
model includes the bulge, bar, exponential disc, and halo. We used
the analytical Ferrers ellipsoid to calculate the potential of the
bar \citep{freeman1972, pfenniger1984, binney2008, sellwood1993}. The
major and minor axes of the bar are $a=3.5$ and $b=1.35$ kpc,
respectively. The mass of the bar is $1.2 \times 10^{10}$ M$_\odot$.
The angular velocity of the bar rotation is $\Omega_b=55$ km s$^{-1}$
kpc$^{-1}$ which corresponds to the best agreement between the model
and observed distributions of the velocities $V_R$ and $V_T$ in
Galactocentric distance $R$. The CR of the bar is located at
$R_{RC}=4.04$ kpc, the resonance $-4/1$ and the OLR of the bar lie at
$R_{-4/1}=5.52$ and $R_{OLR}=7.00$ kpc, respectively. The bar grows
slowly, approaching full strength by $T_g\sim 0.45$ Gyr, which is
equal to four bar rotation periods, but the $m=0$ component of the
bar  is included in the model from the very beginning. It means that
the model is initially axisymmetric, but then its non-axisymmetric
perturbation, i.e. the bar component, slowly increases. The mass of
the bar is conserved during the simulation. The strength of the bar
after it reaches a full power is $Q_b=0.3142$, which corresponds to
strong bars \citep{block2001, buta2004, diaz-garcia2016}.

The model of the Galaxy includes the exponential disc with the mass
of $M_d=3.25 \times 10^{10}\,$M$_\odot$ and the characteristic scale
of $R_d=2.5$ kpc. The total  mass of the model disc and the bar is
$4.45 \times 10^{10}\,$M$_\odot$, which agrees with other estimates
of the mass of the Galactic disc lying in range 3.5--5.0$\times
10^{10}$M$_\odot$ \citep{shen2010, fujii2019}.

The average azimuthal velocities $\overline{V_T}$ in the velocity
distribution of stars at $t=0$ are determined   by solving the Jeans
equation. We set the radial scale length of the initial distribution
of radial velocity dispersion equal to $20$ kpc.

The classical bulge is represented by a Plummer sphere with a mass of
$M_{bg}=5\times 10^{9}\,$M$_\odot$ \citep{nataf2017, fujii2019}. The
halo is modeled as an isothermal sphere \citep{binney2008}.

Model particles represent the motions of  stars of the thin disc. The
initial  radial-velocity dispersion of model particles at the solar
distance is $\sigma_R=30.5$ km s$^{-1}$, which is close to the
observed value, $\sigma_R=32.0$ km s$^{-1}$ ({\it Gaia} DR3).

Note that our model has an order of symmetry equal to $m=2$, i.e.
each star with coordinates ($x$, $y$)  and velocities ($V_x$, $V_y$)
has a twin with coordinates ($-x$, $-y$) and velocities ($-V_x$,
$-V_y$). For a more detailed description of the model, see
\citet{melnik2021}.

\section{Periodic changes in the ring morphology}
\begin{figure}
\resizebox{\hsize}{!}{\includegraphics{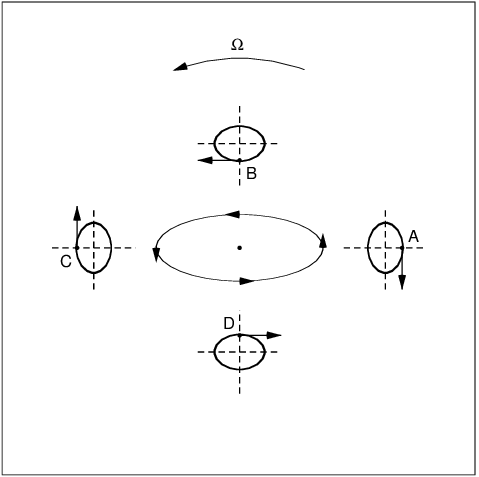}} \caption{Motion of
a star in an elliptical orbit supporting the bar. The star is located
inside the CR and its motion is considered in the reference frame of
the rotating bar. On the major axis of the bar, the star is at the
outermost points of the epicycle ($A$ and $C$) and the positive
velocity of  orbital rotation (i.e. motion in the sense of Galactic
rotation)  adds up with the negative velocity of the epicyclic
motion, and vice versa, on the minor axis of the bar, the star is at
the points of the epicycle closest to  the center ($B$ and $D$) and
the positive velocity of the orbital rotation  adds up with the
positive velocity of the epicyclic motion. This means that the star
moves at a smaller velocity on the major axis of the bar than on the
minor one. Consequently, the density of stars on the major axis of
the bar must be higher than on the minor one. }
\label{schema_epicycle}
\end{figure}

\subsection{Density distribution in the Galactic disc}

The model stellar disc forms the outer resonance rings by the time
$\sim 1$ Gyr. However, the perturbations of the stellar density in
the disc are small which is a consequence of  the large velocity
dispersion. To see the density perturbations  we should  increase the
density contrast in the disc.

Fig.~\ref{distrib} shows the distributions of stars in the Galactic
plane at the time instants 0.5, 1.0, 1.5, 2.0, 2.5, and 3.0 Gyr
processed by a program that increases the contrast. The parameter $h$
controls the contrast: the higher $h$, the higher the contrast. The
essence of processing procedure is that for each star we calculate
the average density $\Sigma_1$ of other stars located within a radius
of 50 pc from the star considered and compare $\Sigma_1$ with the
initial density $\Sigma_0$ at $t=0$, when the distribution of stars
is purely exponential. If $\Sigma_1/\Sigma_0\le 1.0$ then the star is
not included in the image. If $\Sigma_1/\Sigma_0\ge h$, then the star
is included in the image with the probability $P=100$ per cent. If
the density ratio has intermediate values, $1.0<\Sigma_1/\Sigma_0<h$,
then the star is included in the image with the probability:

\begin{equation}
P=\frac{\Sigma_1-\Sigma_0}{\Sigma_0 (h-1)}\, 100\%,
 \label{probab}
\end{equation}

\noindent which increases linearly from 0 to 100 per cent with
increasing the ratio $\Sigma_1/\Sigma_0$ from 1.0 to $h$.  The
contrast parameter in Fig.~\ref{distrib} is $h=1.5$.

Fig.~\ref{distrib} shows the presence of an inner pseudo-ring located
between the CR and the resonance $-4/1$, the outer resonance ring
$R_2$ located outside the OLR ($R> R_{OLR}$), as well as the
appearance of either trailing ($t=0.5$, 1.0, 2.5, 3.0 Gyr) or leading
($t=1.5$, 2.0 Gyr) spiral arms located in the region between the
resonance $-4/1$ and the OLR which are  the trailing  and leading
segments of the ring $R_1$, respectively. Near the OLR of the bar the
outer ring $R_1$ smoothly transits into the ring $R_2$.

Fig.~\ref{distrib} shows that inside the bar (red ellipse) the
stellar density  increases only near the ends of the bar which  may
come as a surprise.  Though epicyclic approximation is not valid in
the bar region for quantitative estimates, it  can help understand
this fact at a qualitative level. Fig.~\ref{schema_epicycle}
schematically  illustrates the motion of a star in an elliptical
orbit supporting the bar.   We can see that stars move with a smaller
velocity on the major axis of the bar than on the minor one.
Consequently, the density of stars on the major axis of the bar must
be higher than on the minor one. This relation is a consequence of
the motion of stars in closed orbits and the formation of elliptical
orbits from circular ones uniformly populated by stars.

Fig.~\ref{distrib} also shows that the central region of the Galaxy
is  relatively empty. This fact is due to a fast fall of stars onto
the Galactic center during the bar formation. By the time $t=0.5$
Gyr, 19 per cent of stars initially located within the distance
interval $0.02<R<1$ kpc are found to be inside a tiny region of
$R<0.02$ kpc and the integration of their orbits is stopped.  The
rapid loss of the angular momentum  is possibly due to the growth of
the bar rotating at a constant  angular velocity, which is
considerably smaller than the initial angular velocities of most of
stars  in the central region.

In general, we had a choice:  either to compare the local value
(local in place and in time) of the stellar density with the initial
density distribution or with the current averaged value of the
density at the radius considered. The first approach is more suitable
to search for ring structures and the second one -- for
identification of open (with a large pitch angle) spiral arms
\citep[for example,][]{melnik2013}.

An elementary logarithmic spiral wave is defined by the equation:

\begin{equation}
R= C e^{\tan(\gamma) \,(\theta-\theta_1)},
\end{equation}

\noindent where $\gamma$ is the pitch angle of spiral arms,  $R$ and
$C$ are the Galactocentric  distances to the spiral arm at the
Galactocentric angles $\theta$ and $\theta_1$, respectively. If
$\theta$ increases in the sense of  Galactic rotation then the
negative and positive values of $\gamma$ describe the trailing and
leading spiral arms, respectively. Note that if we subdivide an
elliptical ring into four parts along its two axes of symmetry, then
each quarter of the ring can be approximated by a fragment of a
leading or a trailing spiral arm. Thus an  elliptical ring can be
represented by the superposition of  trailing and leading spiral
perturbations with the same absolute values of the pitch angle
$|\gamma|$.

The amplitudes of the spiral oscillations in the distribution of N
objects in the galactic plane can be determined from the formula:

\begin{equation}
A(p,m)=\frac{1}{N} \sum_{j=1}^N e^{-i(m\theta_j+p\ln R_j)},
\end{equation}

\noindent where $p=-m/\tan \gamma$ and $m$ is the number of spiral
arms \citep{kalnajs1971, considere1982, binney2008}. As elliptical
rings have an order of symmetry equal to $m=2$, hereafter we consider
only  $m=2$ amplitudes. The parameter $p$ determines the pitch angle
$\gamma$ of spiral arms: $p>0$  and $p<0$ correspond to trailing and
leading spiral arms, respectively.

\begin{figure*}
\resizebox{\hsize}{!}{\includegraphics{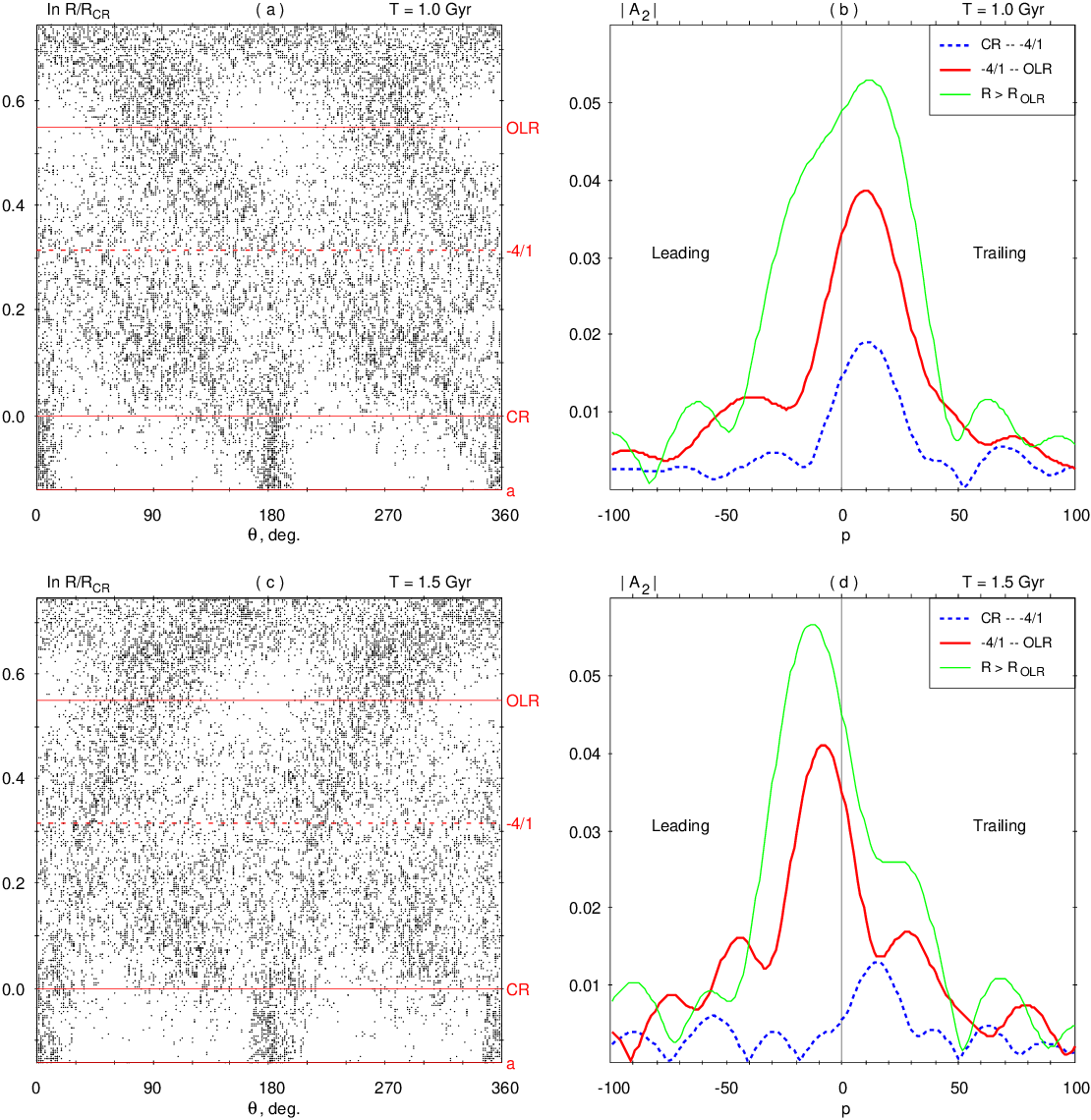}} \caption{Positions
of the spiral arms and the Fourier amplitudes $|A_2|$ calculated for
the time instants $t=1.0$ Gyr (the upper row) and $t=1.5$ Gyr (the
bottom row). (a, c) Distributions of stars in the  ($\theta$, $\ln
R/R_{CR}$) plane. The red  horizontal lines indicate the positions of
the resonances and  the bar major axis $a$. We can see that stars
concentrate to the straight lines that correspond to the positions of
the logarithmic spiral arms. The slope of the lines  to the left
($t=1.0$ Gyr) or to the right ($t=1.5$ Gyr) indicates the
predominance of trailing ($t=1.0$ Gyr) or leading ($t=1.5$ Gyr)
spiral arms, respectively. (b, d) The Fourier amplitudes $|A_2|$ of
the expansions of the distributions of stars into spiral harmonics
calculated in three regions: $R_{CR}<R<R_{-4/1}$ (inner ring),
$R_{-4/1}<R<R_{OLR}$ (outer ring $R_1$) and $R_{OLR}<R<R_{OLR}+1.5$
kpc (outer ring $R_2$). The parameter $p$ determines the pitch angle
$\gamma$ of spiral arms, $p=-m/\tan \gamma$, where $p>0$ and $p<0$
correspond to  trailing and leading spiral arms, respectively. The
vertical line corresponds to $p=0$. We can see that at the time
$t=1.0$ Gyr maximum values of the amplitudes $|A_2|$ indicate the
predominance of  trailing spiral arms in all three regions, but at
$t=1.5$ Gyr  trailing arms dominate only in the region of the inner
ring, $R_{CR}<R<R_{-4/1}$, whereas leading spirals prevail in the
other two regions.} \label{ln_four_1}
\end{figure*}

Fig.~\ref{ln_four_1} (a, c) shows the distributions of stars in the
plane ($\theta$, $\ln R/R_{CR}$) computed for the time instants
$t=1.0$ and $t=1.5$ Gyr. It is evident from the figure that stars
concentrate to the straight lines which correspond to the positions
of the logarithmic spiral arms. The slope of the lines to the left
($t=1.0$ Gyr) or to the right ($t=1.5$ Gyr) indicates the
predominance of trailing ($t=1.0$ Gyr) or leading ($t=1.5$ Gyr)
spiral arms.

Fig.~\ref{ln_four_1} (b, d) shows the amplitudes $|A_2|$ of the
Fourier  transforms calculated in three regions: $R_{CR}<R<R_{-4/1}$
(inner ring), $R_{-4/1}<R<R_{OLR}$ (outer ring $R_1$), and
$R_{OLR}<R<R_{OLR}+1.5$ kpc (outer ring $R_2$). We can see that at
the time moment $t=1.0$ Gyr maximum values of the amplitude $|A_2|$
correspond to $p \approx 10$ in all three regions, which indicates
the predominance of   trailing spiral arms with the pitch angle of
$\gamma \approx -10^\circ$. At the time $t=1.5$ Gyr, trailing spiral
arms dominate only in the region of the inner ring,
$R_{CR}<R<R_{-4/1}$, while  leading spirals dominate in two other
regions.

Note that we show the distribution of stars in the plane ($\theta$,
$\ln R/R_{CR}$) (Fig.~\ref{ln_four_1} a, c)  only for stars that
remain in the sample after processing the sample by the contrast
increasing program (Eq.~\ref{probab}), but compute the Fourier
amplitudes $|A_2|$  with the use of all stars located in the region
considered.

\begin{figure*}
\resizebox{\hsize}{!}{\includegraphics{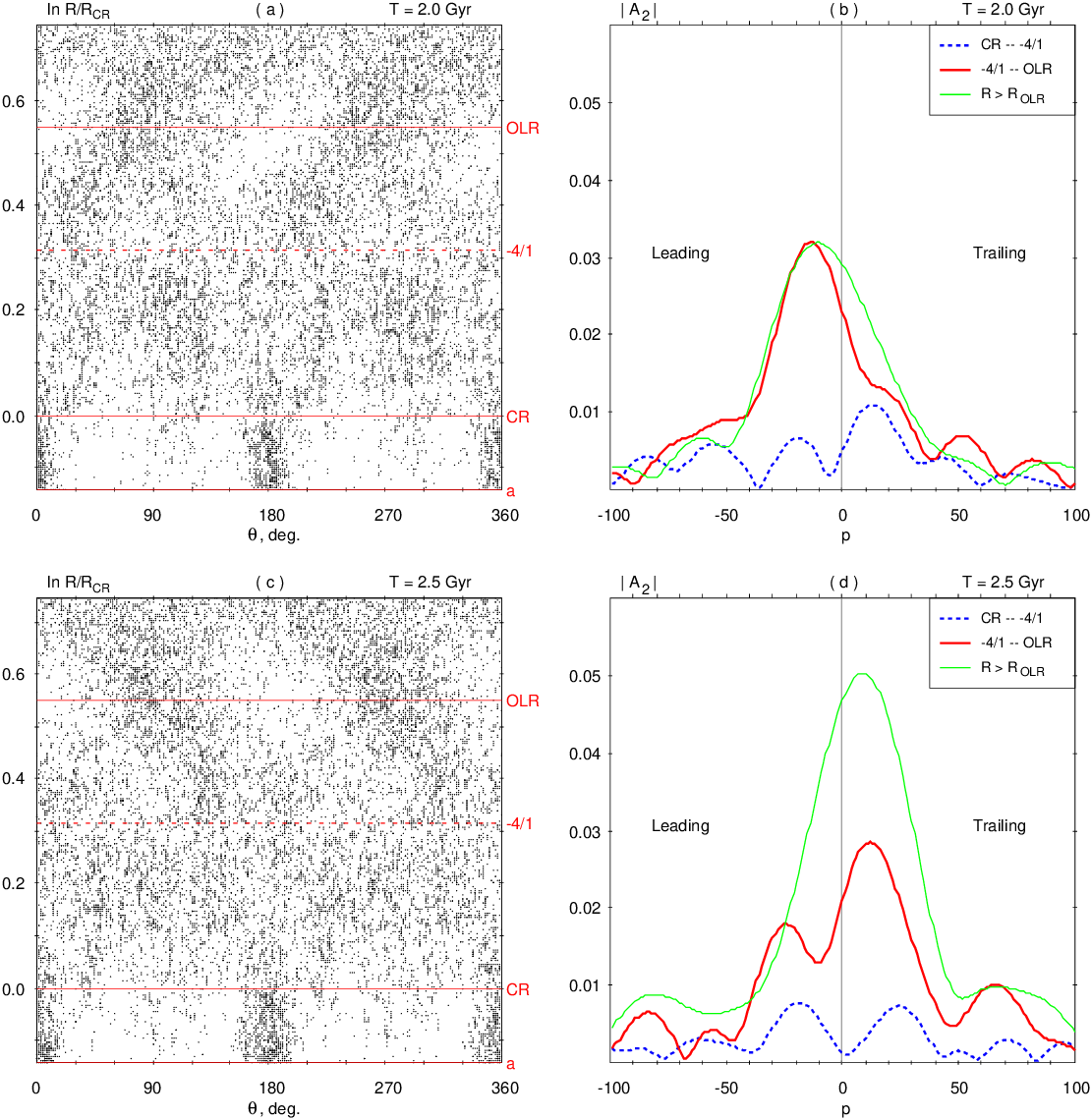}} \caption{Positions
of the spiral arms and the Fourier amplitudes $|A_2|$ calculated for
the time instants $t=2.0$ Gyr (the upper row) and $t=2.5$ Gyr (the
bottom row). (a, c) It can be seen that in the region of the outer
rings, $R_{-4/1}<R<R_{OLR}$+1.5 kpc,  stars concentrate to the
straight lines inclined to the   right ($t=2.0$ Gyr) or to the left
($t=2.5$ Gyr), which indicates the predominance of leading ($t=2.0$
Gyr) or trailing ($t=2.5$ Gyr) spirals, respectively. (b, d) In the
region of the outer rings, maximum values of the amplitude $|A_2|$
correspond to negative ($t=2.0$ Gyr) or  positive ($t=2.5$ Gyr) value
of $p$, which indicates the predominance of leading ($t=2.0$ Gyr) or
trailing ($t=2.5$ Gyr) spiral arms, respectively. In the region of
the inner ring, $R_{CR}<R<R_{-4/1}$, we can see the predominance of
trailing spirals at $t=2.0$ Gyr and the symmetry in the amplitude
$|A_2|$ with respect to the vertical line $p=0$ at the time $t=2.5$
Gyr which is typical for elliptical structures. For more details, see
caption to Fig.~\ref{ln_four_1}.} \label{ln_four_2}
\end{figure*}

Fig.~\ref{ln_four_2} (a, c) shows the distributions of stars in the
plane ($\theta$, $\ln R/R_{CR}$) at $t=2.0$ and $t=2.5$ Gyr. As is
evident from the figure, stars in the region of the outer rings
concentrate to the straight lines inclined to the right ($t=2.0$ Gyr)
and to the left ($t=2.5$ Gyr) which indicates the predominance of
leading ($t=2.0$ Gyr) and trailing ($t=2.5$ Gyr) spiral arms,
respectively.

Fig.~\ref{ln_four_2} (b, d) shows that in the region of the outer
rings maximum values of the amplitudes $|A_2|$ correspond to negative
($t=2.0$ Gyr) or  positive   ($t=2.5$ Gyr) values of $p$,
respectively, which indicates the predominance of leading ($t=2.0$
Gyr) or trailing ($t=2.5$ Gyr) spiral arms. In the region of the
inner ring, $R_{CR}<R<R_{-4/1}$, we see the predominance of trailing
spirals at the  time $t=2.0$ Gyr and the symmetry in the distribution
of the amplitude $|A_2|$ with respect to the vertical line $p=0$ at
the time $T =2.5$ Gyr, which is typical for elliptical structures. A
comparison of Fig.~\ref{ln_four_1} and Fig.~\ref{ln_four_2} reveals
the  changes in maximum values of the amplitudes $|A_2|$ in each
region.

Thus, we have found changes in the morphology of the inner ring and
outer rings, namely the enhancement  (density increase) of either
trailing or leading segments of the elliptical rings.

To prove that the changes in morphology discussed here are not due to
random fluctuations in  density and velocity, we build model 2, which
differs from model 1 (considered here) by random deviations in the
density and velocity distributions at $t=0$. The random changes are
hardly seen in the distribution of stars in the Galactic plane, but
they are visible in  variations  of  the Fourier amplitudes
(Fig.~\ref{amp_p}) and we discuss them in section 4.2.


\subsection{Periodic changes in the
density of trailing and leading segments of the resonance rings}

\begin{figure*}
\resizebox{\hsize}{!}{\includegraphics{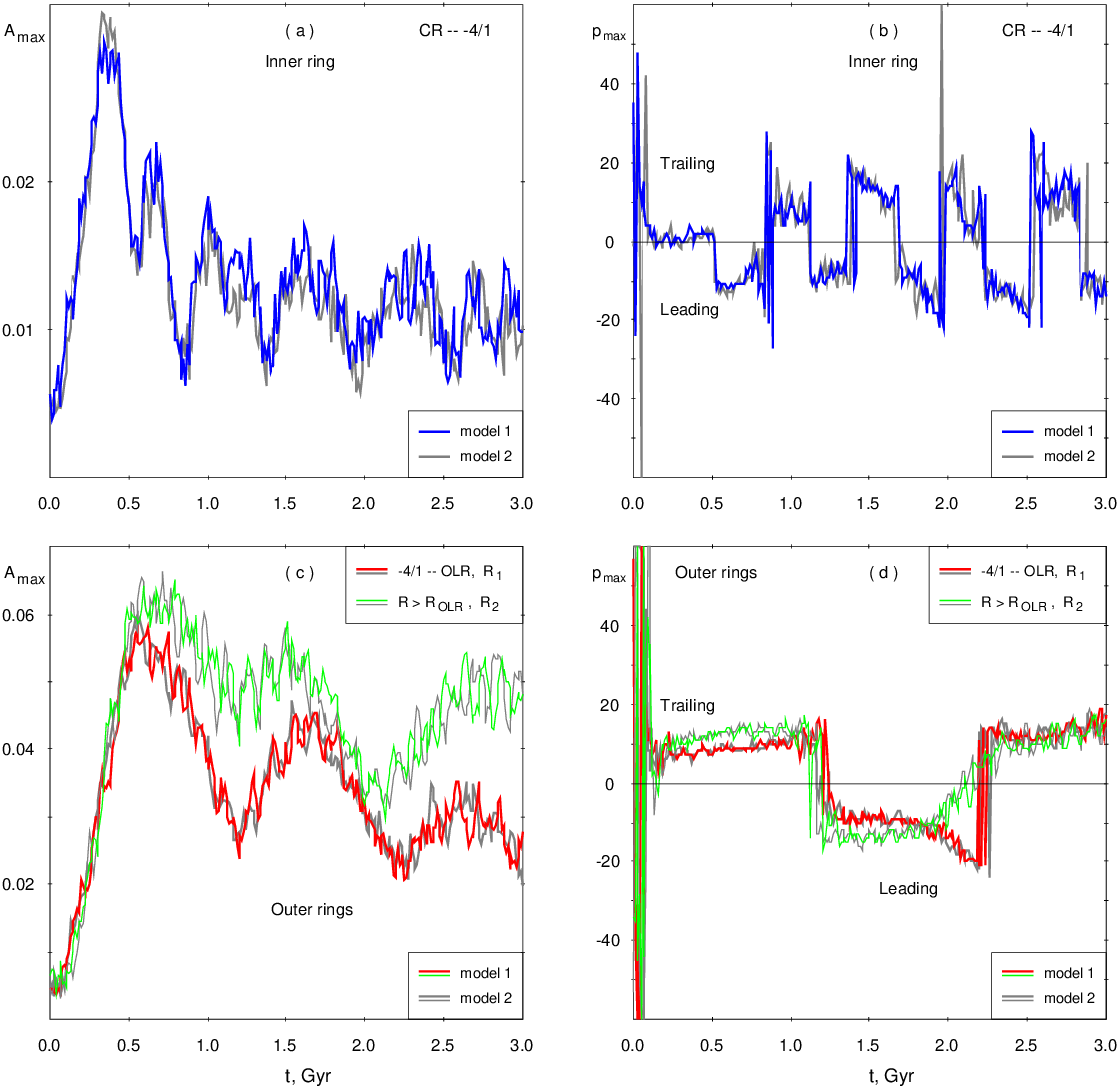}} \caption{(a, c)
Variations in  maximum values, $A_{max}$, of  the Fourier amplitude
$|A_2|$ calculated in the region of the inner ring (a) and  the outer
rings (c) as a function of time $t$. (b, d) Variations of the
parameter $p_{max}=-m/\tan\gamma$ corresponding to maximum value of
the amplitude $|A_2|$ computed in the region of the inner ring (b)
and in the region of the  outer rings (d) as a function of time.  We
can see that the parameters calculated for models 1 and 2 agree
within the random fluctuations. The period of changes in the
amplitude $A_{max}$ in the regions of the inner ring (a) is much
shorter than that in the region of the outer rings (c). Variations in
$A_{max}$ and $p_{max}$ calculated in the region of the outer ring
$R_1$ ($R_{-4/1}<R<R_{OLR}$) and in the region of the outer ring
$R_2$ ($R_{OLR}<R<R_{OLR}+1.5$ kpc) are practically synchronized.}
\label{amp_p}
\end{figure*}

Let us consider periodic changes in the morphology of the resonance
rings and determine the periods of these variations. Fig.~\ref{amp_p}
(left panel) shows the variations in  maximum values, $A_{max}$, of
the Fourier amplitude $|A_2|$ calculated in the region of the inner
ring and in the region of the outer rings as a function of time $t$.
Fig.~\ref{amp_p} (right panel) shows variations in the parameter
$p_{max}$ corresponding to maximum value of the amplitude $|A_2|$.
The parameter $p_{max}=-m/\tan \gamma$ characterizes the pitch angle
$\gamma$ of a two-armed spiral pattern, which predominates at the
time considered.  We can see that  the parameters calculated for
models 1 and 2 that differ from one another by random deviations of
their initial conditions agree within the random fluctuations. The
period of variations of the amplitude $A_{max}$ in the region of the
inner ring ($R_{CR}<R<R_{-4/1}$) is significantly shorter than that
in the region of the outer rings $R_1$ and $R_2$
($R_{-4/1}<R<R_{OLR}$ and $R_{OLR}<R<R_{OLR}+1.5$ kpc, respectively).
The same is true for variations of the parameter $p_{max}$: the
period of changes of $p_{max}$ in the region of the inner ring is
much shorter than  in the region of the outer rings. Changes in the
amplitude $A_{max}$ in the region of the outer ring $R_1$
($R_{-4/1}<R<R_{OLR}$) and in that of the outer ring $R_2$
($R_{OLR}<R<R_{OLR}+1.5$ kpc) are practically synchronized.

To a first approximation, the changes in the amplitude $A_{max}$ and
in the parameter $p_{max}$ can be represented as cosine oscillations
with a period $P$ about the average value:

\begin{equation}
A_{max}=C_0+C_1\sin(2\pi t/P)+C_2\cos(2\pi t/P), \label{osc_1}
\end{equation}

\begin{equation}
p_{max}=D_0+D_1\sin(2\pi t/P)+D_2\cos(2\pi t/P), \label{osc_2}
\end{equation}

\noindent where $C_0$ and $D_0$ are average values,
$C_3=\sqrt{C_1^2+C_2^2}$ and $D_3=\sqrt{D_1^2+D_2^2}$ are the
amplitudes of oscillation, but $\phi_a=\arctan(C_2/C_1)$ and
$\phi_p=\arctan(D_2/D_1)$ are the  initial phases of oscillations of
$A_{max}$ and $p_{max}$, respectively.

We consider oscillations only starting from the time instant $t=0.5$
Gyr when the bar reaches its full strength.
Eqs.~\ref{osc_1}~and~\ref{osc_2} are linear with respect to the
parameters $C_0$, $C_1$, $C_2$, $D_0$, $D_1$, $D_2$ and non-linear
with respect to  the oscillation period $P$. We have considered
values of $P$ in the range  $P\in [0.03, 5.0]$ Gyr with a step of
0.01 Gyr. For each value of $P$, the linear coefficients and the
$\chi^2$ function (the sum of squared normalized differences between
the analytic representation, Eq.~\ref{osc_1}--\ref{osc_2}, and the
observed values, Fig.~\ref{amp_p}) are calculated.

\begin{figure*}
\resizebox{\hsize}{!}{\includegraphics{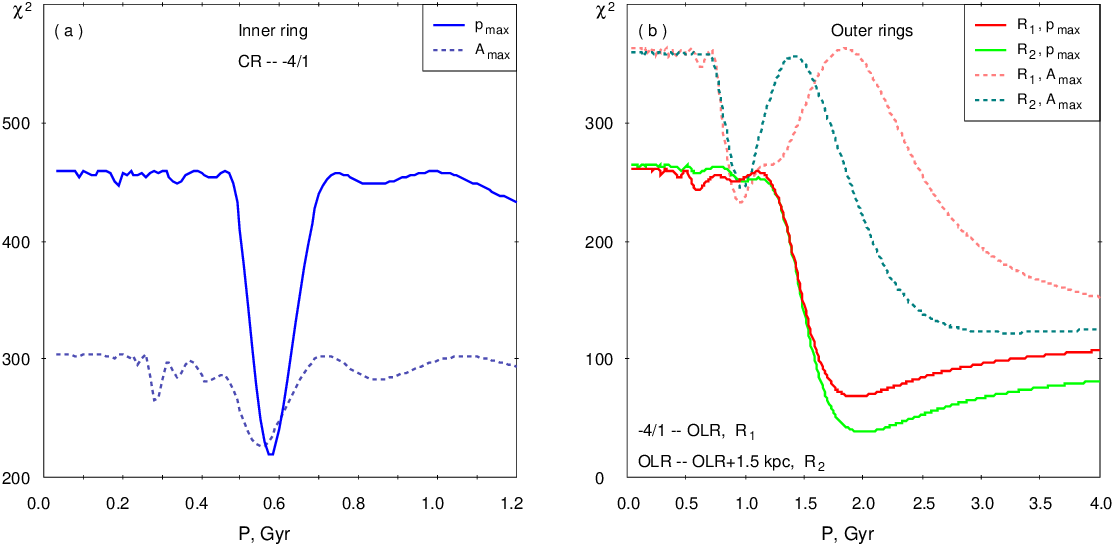}}
\caption{Dependences of the $\chi^2$ values on the period $P$
calculated for the parameter $p_{max}=-m/\tan \gamma$ (solid lines)
and for the amplitude $A_{max}$ (dashed lines)  in model 1. Functions
$\chi^2$ are calculated (a) in the region of the inner ring
($R_{CR}<R<R_{-4/1}$) and (b) in the regions of the outer rings $R_1$
and $R_2$ ($R_{-4/1}<R<R_{OLR}$ and $R_{OLR}<R<R_{OLR}+1.5$ kpc,
respectively). (a) In the region of the inner ring,  the $\chi^2$
functions derived from the analysis of oscillations of the parameter
$p_{max}$ (solid line) and of the amplitude $A_{max}$ (dashed line)
reach their minimum values at $P=0.58\pm0.02$ and $0.55\pm0.02$ Gyr,
respectively. (b) In the region of the outer rings, the $\chi^2$
functions derived from the analysis of oscillations of the parameter
$p_{max}$ (solid lines) reach their minimum values  at
$P=1.93\pm0.09$ Gyr ($R_1$) and $P=2.00\pm0.09$ Gyr ($R_2$); the
$\chi^2$ functions calculated from the analysis of the amplitude
$A_{max}$ (dashed lines) reach the local minima at $P=0.96\pm0.02$
Gyr ($R_1$) and $P=0.97\pm0.02$ Gyr ($R_2$) followed by the local
maxima and a sharp drop at $P>2.0$ Gyr.} \label{chi_per}
\end{figure*}

Fig.~\ref{chi_per} (a) shows the $\chi^2$ functions computed  for
model 1 in the region of the inner ring ($R_{CR}<R<R_{-4/1}$). The
$\chi^2$ functions calculated from the analysis of oscillations of
the parameter $p_{max}$ (solid line) and  of the amplitude
$A_{max}$(dashed line) reach minimum values at $P=0.58\pm0.02$ and
$0.55\pm0.02$ Gyr, respectively. In the region of the inner ring the
estimates  of the period  derived from the time series of $A_{max}$
and  $p_{max}$ can be seen to agree well with each other.

The  periods obtained for model 2 in the region of the inner ring
derived from variations of  $p_{max}$ and  $A_{max}$ are
$0.58\pm0.02$ and  $P=0.54\pm0.02$ Gyr, respectively.

Fig.~\ref{chi_per}~(b) shows the $\chi^2$ functions computed  for
model 1 in the regions of the outer ring $R_1$ ($R_{-4/1}<R<R_{OLR}$)
and the outer ring $R_2$ ($R_{OLR}<R<R_{OLR}+1.5$ kpc). The $\chi^2$
functions derived from the analysis of variations of the parameter
$p_{max}$ (solid lines) reach minimum values at $P=1.93\pm0.09$
($R_1$) and $P=2.00\pm0.09$ ($R_2$) Gyr. The $\chi^2$ functions
calculated from the analysis of oscillations of the amplitude
$A_{max}$ (dashed lines) reach the local minimum at $P=0.96\pm0.02$
Gyr ($R_1$) and $P=0.97\pm0.02$ Gyr ($R_2$) followed by the  local
maximum and a sharp drop at $P>2.0$ Gyr.

The  periods obtained for model 2 in the regions of the outer rings
derived from variations of $p_{max}$ and $A_{max}$  have following
values: $P=1.97\pm0.09$ ($R_1$, $p_{max}$), $2.10\pm0.09$ ($R_2$,
$p_{max}$), $0.95\pm0.02$ ($R_1$, $A_{max}$), and $1.00\pm0.02$ Gyr
($R_2$, $A_{max}$).

Thus,  the corresponding values of the periods obtained for models 1
and 2 that differ by random deviations in the initial conditions
coincide within the random errors. In the region of the inner ring,
the period of oscillations of $p_{max}$ and $A_{max}$ is
$P=0.57\pm0.02$ Gyr while in the region of the outer rings, the
periods of oscillations are $P=2.0\pm0.1$ and $0.97\pm0.02$ Gyr,
respectively.


\section{Orbits supporting the inner combined ring }

\subsection{Periods of oscillations of stars near the equilibrium points $L_4$
and $L_5$}

\begin{figure*}
\resizebox{\hsize}{!}{\includegraphics{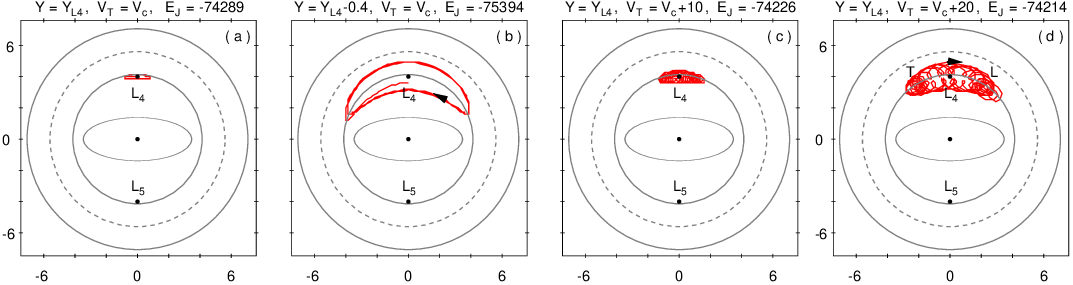}} \caption{Some
examples of orbits near the stable equilibrium point $L_4$ ($L_5$).
All frames show orbits of stars with the initial coordinate $X(0)=0$
(the star is on the minor axis of the bar at $t=0$) and the initial
radial velocity $V_R=0$.  The initial coordinate $Y$ and the initial
azimuthal velocity $V_T$  are indicated on the frames.  The velocity
$V_c$ is the velocity of the rotation curve at the distance of the
point $L_4$.  For each star we present the Jacobi integral, $E_J$,
which is conserved  over the time interval $t>T_g$. The additional
velocities, coordinates, and Jacobi integrals are given in units of
km s$^{-1}$, kpc, and km$^2$ s$^{-2}$, respectively. The position of
the bar is shown by an ellipse. The gray solid lines indicate the CR
and OLR, and the dashed line shows the resonance $-4/1$. Short-period
orbits (SPO) are similar to small ellipses but long-period orbits
(LPO) resemble a banana.   Frame (d) also shows that the outer parts
of banana-shaped orbits located outside the CR, $R>R_{CR}$, can be
represented as a superposition of trailing and leading segments
marked by the letters 'T' and 'L', respectively. The galaxy rotates
counterclockwise. If an overdensity forms at some part of
banana-shaped orbits, then it starts circulating  the closed path
causing the enhancement of either leading or trailing segments of the
inner ring.} \label{orb_L_4}
\end{figure*}

Let us consider orbits near the Lagrange equilibrium points $L_4$ and
$L_5$.  Fig.~\ref{orb_L_4} shows examples of orbits near the stable
equilibrium point $L_4$ ($L_5$). The initial coordinate $X$ of the
star is  $X(0)=0$ (the star is located on the minor axis of the bar)
and the initial radial velocity is $V_R=0$ in all four examples. The
initial coordinate $Y$ and the initial azimuthal velocity are
indicated on the frames.  For each star we also present  the Jacobi
integral, $E_J$ (for more details, see section 6),  which is
conserved over the time interval $t>T_g$ when the bar becomes time
independent. Frame~\ref{orb_L_4} (a) shows the orbit of a star
located at the point $L_4$ at the initial instant with the velocity
equal to the velocity of the rotation curve $V_c$ at the distance of
the point $L_4$. The orbit resembles a small ellipse with the star
moving clockwise   (i.e. in the sense opposite that of galactic
rotation). This  is an example of a short-period orbit.
Frames~\ref{orb_L_4} (b, c, d) show some examples of a long-period
orbit which resemble a banana.

We use the distribution of the potential in the model disc to
calculate the periods of short- and long-period  oscillations near
the equilibrium point $L_4$ ($L_5$) and compare them with the periods
of variations of $A_{max}$ and $p_{max}$ in the region of the inner
ring.

Motions of stars in the reference frame of  the bar rotating with the
angular velocity $\Omega_b$ are determined by the following equation:

\begin{equation}
\ddot{\bf x}=-\nabla \Phi_\textrm{eff}-2 {\bf\Omega_b} \times
\dot{\bf x}, \label{ddot_x}
\end{equation}

\noindent where  the coordinates ${\bf x}$ and the velocities
$\dot{\bf x}$ characterize the  motion in the rotating reference
frame, the effective potential $\Phi_\textrm{eff}$ includes both the
gravitational potential and the potential of  the centrifugal force:

\begin{equation}
\Phi_\textrm{eff}=\Phi - \frac{1}{2} \,\Omega_b^2 R^2, \label{phi}
\end{equation}

\noindent and the term $-2{\bf\Omega_b} \times \dot{\bf x}$
characterizes the contribution of the Coriolis force.

Let us consider the motion of stars relative to the equilibrium point
$L_4$ lying on the minor axis of the bar:

\begin{equation}
\xi=x- x_L, \label{xi}
\end{equation}
\begin{equation}
\eta=y- y_L, \label{eta}
\end{equation}

\noindent the coordinates $\xi$ and $\eta$ are measured  along the
major and minor axes of the bar, respectively. Expansion of
$\Phi_\textrm{eff}$ into Taylor series at the equilibrium point $L_4$
and a consideration of  the equilibrium condition:

\begin{equation}
\nabla \Phi_\textrm{eff} (L_4)=0, \label{equilib}
\end{equation}

\noindent give us a possibility to write the equations of motion in
the following form:

\begin{equation}
\ddot{\xi}= 2\Omega_b \dot{\eta} -  \Phi_{xx} \xi, \label{ddot_xi}
\end{equation}
\begin{equation}
\ddot{\eta}= -2\Omega_b \dot{\xi} -  \Phi_{yy} \eta, \label{ddot_eta}
\end{equation}

\noindent where the second derivatives $\Phi_{xx}$ and $\Phi_{xx}$
are taken  at the equilibrium point $L_4$:

\begin{equation}
\Phi_{xx}=\frac{\partial^2 \Phi_\textrm{eff}}{\partial x^2} (L_4),
\label{phi_xx}
\end{equation}
\begin{equation}
\Phi_{yy}=\frac{\partial^2 \Phi_\textrm{eff}}{\partial y^2} (L_4).
\label{phi_xx}
\end{equation}

\noindent Let us assume that a star oscillates about the equilibrium
point $L_4$  with the frequency $\omega$. Then we can write:

\begin{equation}
\xi=\xi_0 \exp(-i\omega t), \label{xi_exp}
\end{equation}
\begin{equation}
\eta=\eta_0 \exp(-i\omega t). \label{eta_exp}
\end{equation}

\noindent Substituting expressions Eq.~\ref{xi_exp}--\ref{eta_exp} in
the system of differential equations
Eq.~\ref{ddot_xi}--\ref{ddot_eta}, we obtain a set of linear
equations that has a non-trivial solution if its determinant is zero:

\begin{equation}
\omega^4-(\Phi_{xx}+\Phi_{yy}+4 \Omega_b^2)\;\omega^2 +
\Phi_{xx}\Phi_{yy}=0. \label{charact}
\end{equation}

\noindent The solution of this quadratic equation with respect to
variable $\omega^2$ has two different roots that characterize the
frequency of motion of stars on long- and short-period orbits near
the equilibrium point $L_4$ \citep[for more details, see][p.
178]{binney2008}.

Table~1 lists the coordinates of the point $L_4$ calculated from the
equilibrium condition Eq.~\ref{equilib}, the frequencies $\omega_1$
and $\omega_2$, periods of motions on short- and long-period orbits,
and their uncertainties, respectively.

The frequency of motions in short-period orbits practically coincides
with the  epicyclic frequency at the distance of the equilibrium
point $L_4$, which is  equal to $\kappa=0.0809$ Myr$^{-1}$.

Note that the  rotation period for long-period orbits, $P=565\pm2$
Myr (Table~1), agrees within the errors with the period found from
the analysis of oscillations of $p_{max}$ and $A_{max}$ in the region
of the inner ring, $P=570\pm20$ Gyr (section 4.2).

A possible explanation of the periodic enhancement of either leading
or trailing segments of the inner ring is the presence of a mechanism
which creates an overdensity in certain parts of banana-shaped orbits
which then begins to circulate along a closed path, producing a
density increase on either leading or trailing segments of
banana-shaped orbits. Fig.~\ref{orb_L_4} (d) illustrates such a
possibility. We can see that the outer parts of banana-shaped orbits
located outside the CR, $R>R_{CR}$, can be represented as a
superposition of  trailing  and leading segments.

In general, we can even point the location  where the overdensity
forms. Fig.~\ref{amp_p} (b) shows that  oscillations start with the
enhancement of leading segments of the inner ring which occurs during
the time interval  $T \sim 0.52$--0.89 Gyr, i.e. just after the bar
gains its full strength ($T_g=0.45$ Gyr). This means that at the time
of the bar formation, the overdensity must be located at the top
point of banana-shaped orbits, i.e. at the point farthest from the
Galactic center. Then the overdensity begins to move clockwise which
causes, in the first turn, the enhancement of leading segments of the
inner ring.

\begin{figure*}
\resizebox{\hsize}{!}{\includegraphics{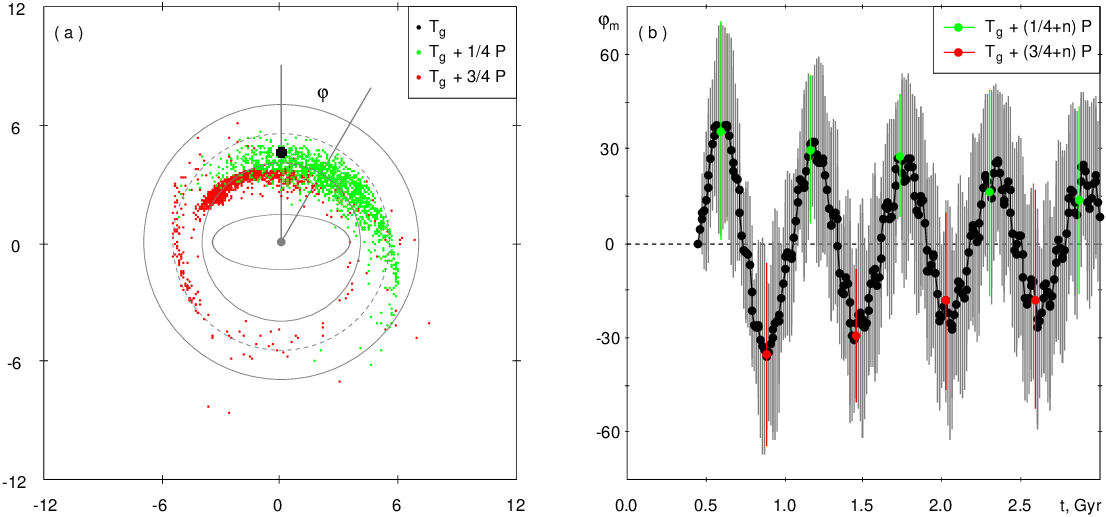}} \caption{ (a)
Locations of the selected stars at  different time instants. The
selected stars are shown in the black, green, and red at the instants
$T_g$, $T_g+1/4 P$, and $T_g+3/4 P$, respectively, where $P$ is close
to the period of  long-period oscillations, $P=0.57$ Gyr. We can see
that most of the selected stars are shifted to the right and to the
left with respect to the minor axis of the bar at the instants
$T_g+1/4 P$ and $T_g+3/4 P$, respectively. The angle $\varphi$ is
measured from to the minor axis of the bar in the sense opposite that
of the Galactic rotation. The position of the bar is shown by an
ellipse. The gray solid lines indicate the CR and OLR, and the dashed
line -- the resonance $-4/1$. (b) Oscillations of the median angle
$\varphi_m$ are computed for the selected stars as a function of
time.   The vertical lines indicate the  $+/-$ dispersion
$\sigma_\varphi$. The green and red colours correspond to the
instants $t=T_g+(1/4+n)P$ and $t=T_g+(3/4+n)P$, where n=0...5, when
the overdensity must achieve the extreme right and left positions,
respectively. We can see that the majority of the selected stars
demonstrates oscillations relative to the minor axis of the bar with
the period of $P=0.57$ Gyr. } \label{overden}
\end{figure*}

To check this hypothesis, we selected stars located within a small
circle  at the time instant $T_g=0.45$ Gyr and traced their motion.
Fig.~\ref{overden} (a) shows the locations of the 6405 selected stars
in the reference frame corotating with the bar at different time
instants. At time $T_g$, the selected stars lie within the circle of
the 200-pc radius  with the center located on the minor axis of the
bar at the distance of $R=R_{CR}+0.5$ kpc from the Galactic center
(black points). The circular shape of the overdensity at the initial
time instant is chosen only for simplicity. After 1/4 of the period
$P=0.57$ Gyr,  which determines the periodic changes of the
morphology of the inner ring (section 4.2), the majority of the
selected stars is shifted to the right with respect to the minor axis
of the bar (green points). However, after $3/4 P$ from the instant
$T_g$, the bulk of the selected stars is shifted to the left (red
points). These displacements can be described with the use of the
angle $\varphi$ measured from the minor axis of the bar in the sense
opposite that of the Galactic rotation, $\varphi=\pi/2-\theta$. We
calculated the median  angle, $\varphi_m$, and  median dispersion,
$\sigma_\varphi$, (half of the central interval containing 68 per
cent of objects)  for the selected stars at different instants.

Note that some of the selected stars appear not to be trapped by the
banana-shaped orbits and continue their rotation around the Galactic
center.

Fig.~\ref{overden} (b) shows the oscillations of the median angle
$\varphi_m$ computed for the selected stars as a function of  time.
The angle $\varphi_m$ varies in the range [$-36$, +38$^\circ$], where
the signs plus and minus correspond to the displacements to the right
and to the left, respectively. Note that the amplitude of the
oscillations of $\varphi_m$ slowly decreases with time.  The vertical
lines show the $+/-$ dispersion $\sigma_\varphi$, with the  average
$\overline{\sigma_\varphi}=27^\circ$. Due to large number of stars in
the overdensity, the average uncertainty in determination of
$\varphi_m$ is only $\varepsilon_{\varphi}=0.4^\circ$. The green and
red colours show the displacements for the instants $t=T_g+(1/4+n)P$
and $t=T_g+(3/4+n)P$ (n=0,..5) when the bulk of the selected stars
must achieve the extreme right and left positions, respectively. We
can see that most of  selected stars show oscillations relative to
the minor axis of the bar and  the period of $P=0.57\pm0.2$ Gyr
describes these oscillations fairly well.

All these results support the hypothesis that the periodic changes in
the morphology of the inner ring are due to the  overdensity forming
at the top point of banana-shaped orbits at the time $T_g=0.45$ Gyr,
when  the bar getting its full strength.

\begin{table}  \caption{Orbit periods near  Lagrange points}
\centering
  \begin{tabular}{lc}
 \\[-7pt]\hline\\[-7pt]
 \multicolumn{2}{c}{L$_4$ (L$_5$)}\\
 $x_{L_4}$ & 0 \\
 $y_{L_4}$ & $3.950\pm0.001$ kpc \\
$\omega_1$ & $0.0803\pm0.0001$ Myr$^{-1}$ \\
$\omega_2$ & $0.01111\pm0.00003$ Myr$^{-1}$ \\
$P_1$ & $78.2\pm0.1$ Myr \\
$P_2$ & $565\pm2$ Myr \\
\\[-7pt]\hline\\[-7pt]
 \multicolumn{2}{c}{L$_1$ (L$_2$)}\\
 $x_{L_1}$ & $4.151\pm0.001$ kpc \\
 $y_{L_1}$ & 0 \\
$\omega_1$ & $0.0794\pm0.0001$  Myr$^{-1}$ \\
$P_1$ & $79.1\pm0.1$ Myr \\
\\[-7pt]\hline\\[-7pt]
\end{tabular}
\label{general}
\end{table}


\subsection{Orbits associated with the equilibrium points $L_1$ and $L_2$ and
their contribution to the formation of the inner ring}

Here we study  the contribution of orbits associated with the
unstable equilibrium points $L_1$ and $L_2$  to the formation of the
inner ring.

We have found  that the shape of  stellar orbits  which start their
motion at  the equilibrium points $L_1$ and $L_2$ depends on the way
of the bar turning on: whether the bar exists from the very beginning
or grows up slowly transforming from a round  into elliptical
structure in four bar rotation periods. Note that during the growth,
the average radial  force created by the bar conserves at any radius
\citep{melnik2021}.

\begin{figure*}
\resizebox{\hsize}{!}{\includegraphics{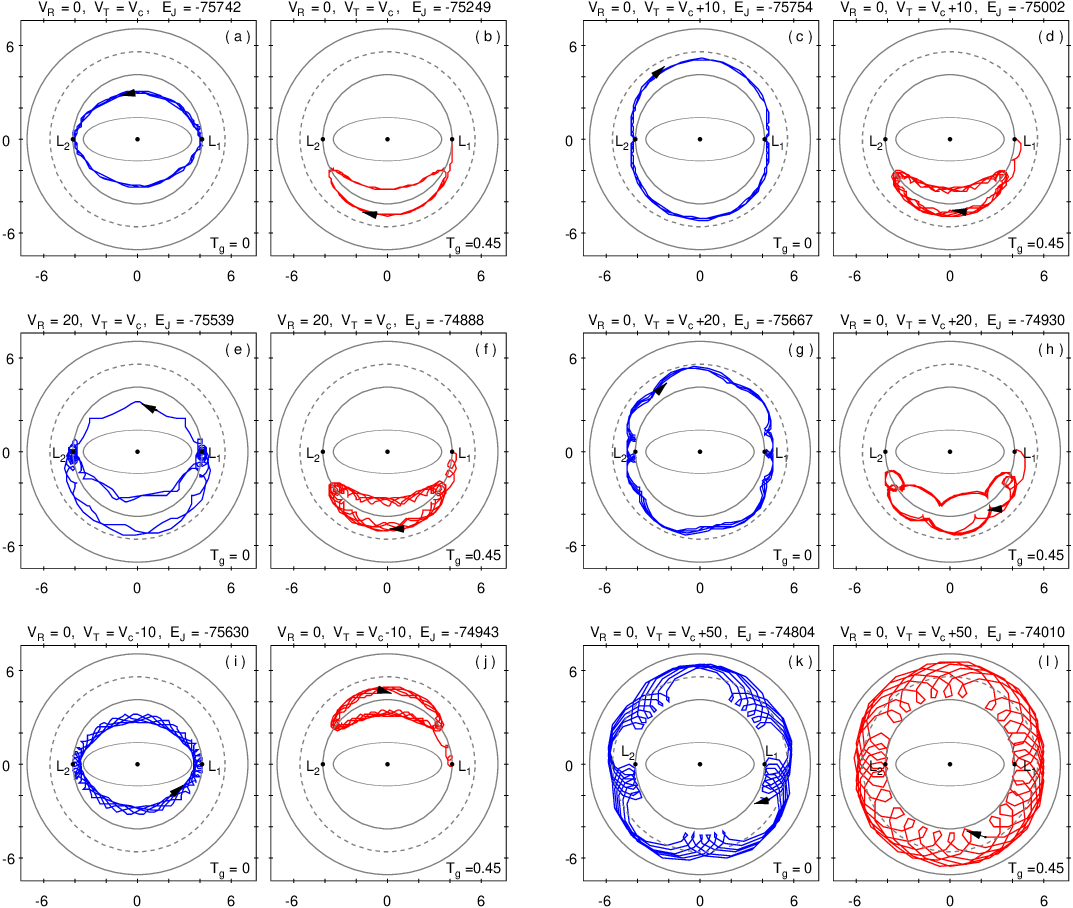}} \caption{Stellar
orbits associated with the equilibrium points $L_1$ and $L_2$
obtained for the potential with  non-axisymmetric perturbations of
the bar existing from the beginning of the simulation (blue orbits)
and for the potential with slowly  growing bar (red orbits). In the
first case the growth time of the bar is $T_g=0$ while in the second
case  $T_g=0.45$ Gyr. The initial positions of all the stars coincide
with the position of the Lagrange equilibrium point $L_1$. The
initial velocities of the stars are indicated on the frames, where
$V_c$ is the velocity of the rotation curve at the distance $L_1$.
For each star we present the Jacobi integral, $E_J$, which is
conserved over the time interval $t>T_g$.  The additional velocities,
coordinates, and Jacobi integrals are given in units of km s$^{-1}$,
kpc, and km$^2$ s$^{-2}$, respectively. The size of the frames is
$15\times 15$ kpc$^2$. Two adjacent frames correspond to the same
initial conditions. Orbits are shown in the reference frame of the
rotating bar, therefore inside the CR stars move in the sense of the
galactic rotation (counterclockwise) but outside the CR -- in the
opposite sense (clockwise). The gray solid lines indicate the CR and
OLR, and the dashed line shows the resonance $-4/1$. We can see that
the orbits in the shape of 'lemon' (a, i), 'eight' (c), 'basket' (e),
and 'flower' (g) obtained for the bar turning on instantly transform
into banana-shaped orbits (b, d, f, h, j) for the slowly growing bar.
The shape of 'lace' orbits (k, l) depends little on the way the bar
turns on. Both banana-shaped and 'lace' orbits support the inner
ring. In addition, 'lace' orbits also support the outer ring $R_1$. }
\label{orb_L1_all}
\end{figure*}

Fig.~\ref{orb_L1_all} shows stellar orbits associated with the
equilibrium points $L_1$ and $L_2$  obtained for the potential with
non-axisymmetric perturbations of the bar existing from  the very
beginning (blue lines) and for the slowly growing bar (red lines).
The bar growth time is $T_g=0$ for the bar turning on instantly and
$T_g=0.45$ Gyr for the slowly growing bar. The initial positions of
all the stars considered coincide with the Lagrange equilibrium point
$L_1$. The initial velocities of the stars are indicated on the
frames. The initial conditions are the same for the two adjacent
frames. For each star we present the Jacobi integral, $E_J$, which is
conserved over the time interval $t>T_g$. During the bar growth the
value of the Jacobi integral changes, namely, it increases.

Fig.~\ref{orb_L1_all} shows that  orbits in the shape of 'lemon' (a,
i), 'eight' (c), 'basket' (e) and 'flower' (g) obtained for the case
of instant bar onset  transform into banana-shaped orbits (b, d, f,
h, j) when the bar turns on slowly. The periods of oscillations on
banana-shaped orbits are $P=0.56$--0.64 Gyr, which are close  but not
exactly coincide with that  of  the long-period oscillations,
$P=0.565\pm0.02$ Gyr (Table~1), around the equilibrium point $L_4$.
All orbits shown in Fig.~\ref{orb_L1_all} support the inner ring.

A comparison with other studies of orbits in barred galaxies suggests
that orbits  in Fig.~\ref{orb_L1_all} (a, i) can be quasi-periodic
ones forming around stable lemon-shaped orbits supporting the bar
\citep[][Figs~10 and 11 therein]{patsis2003}. The eight-shaped orbit
in Fig.~\ref{orb_L1_all} (c) is similar to the unstable orbit with
dimples in study by \citet[][Fig.~3c therein]{contopoulos2006}. The
basket-shaped orbit  in Fig.~\ref{orb_L1_all} (e)  is similar to the
chaotic orbits studied by \citet[][Fig.~14 therein]{patsis2010} and
\citet[][Figs.~12--19 therein]{tsigaridi2013}.

Orbits that resemble a  'lace' (Fig.~\ref{orb_L1_all} k, l) are of
special case. Their shape depends little on the way the bar turns on.
Note that  'lace' orbits are elongated in the direction perpendicular
to the bar and partly lie outside the $-4/1$ resonance, and hence
they also support the outer ring $R_1$ elongated perpendicular to the
bar.

Fig.~\ref{orb_L1_all} shows that when the bar turns on slowly a great
variety of orbits associated with the equilibrium points $L_1$ and
$L_2$ transforms into banana-shaped orbits associated with the
equilibrium points  $L_4$  and $L_5$.

A possible  explanation of this behavior can be given in terms of
resonance trapping \citep{chiba2021, chiba_schonrich2021,
chiba_schonrich2022}. To be trapped by the Corotation Resonance, a
star must  lay in the so-called trapped region in the angle-action
coordinates, which is bounded by the separatrix. If the bar changes
its strength or angular velocity, the location of the separatrix also
changes. The points $L_1$ and $L_2$ are always located outside the
trapped region, so, in the case of instant bar onset, the orbit of a
star, which starts its motion at the point $L_1$, always stay outside
the trapped region (Fig.~\ref{orb_L1_all}, left frames). However,
when the bar turns on slowly,  the Jacobi integral of the star
changes and it has a chance to get inside the trapped region and be
captured by the resonance (Fig.~\ref{orb_L1_all}, right frames except
'lace' orbits).

Thus, in the case of slow bar onset, orbits associated with the
unstable equilibrium points $L_1$ and $L_2$ can also support the
inner combined ring.

\section{Libration and precession of orbits near the OLR}

\begin{figure*}
\resizebox{\hsize}{!}{\includegraphics{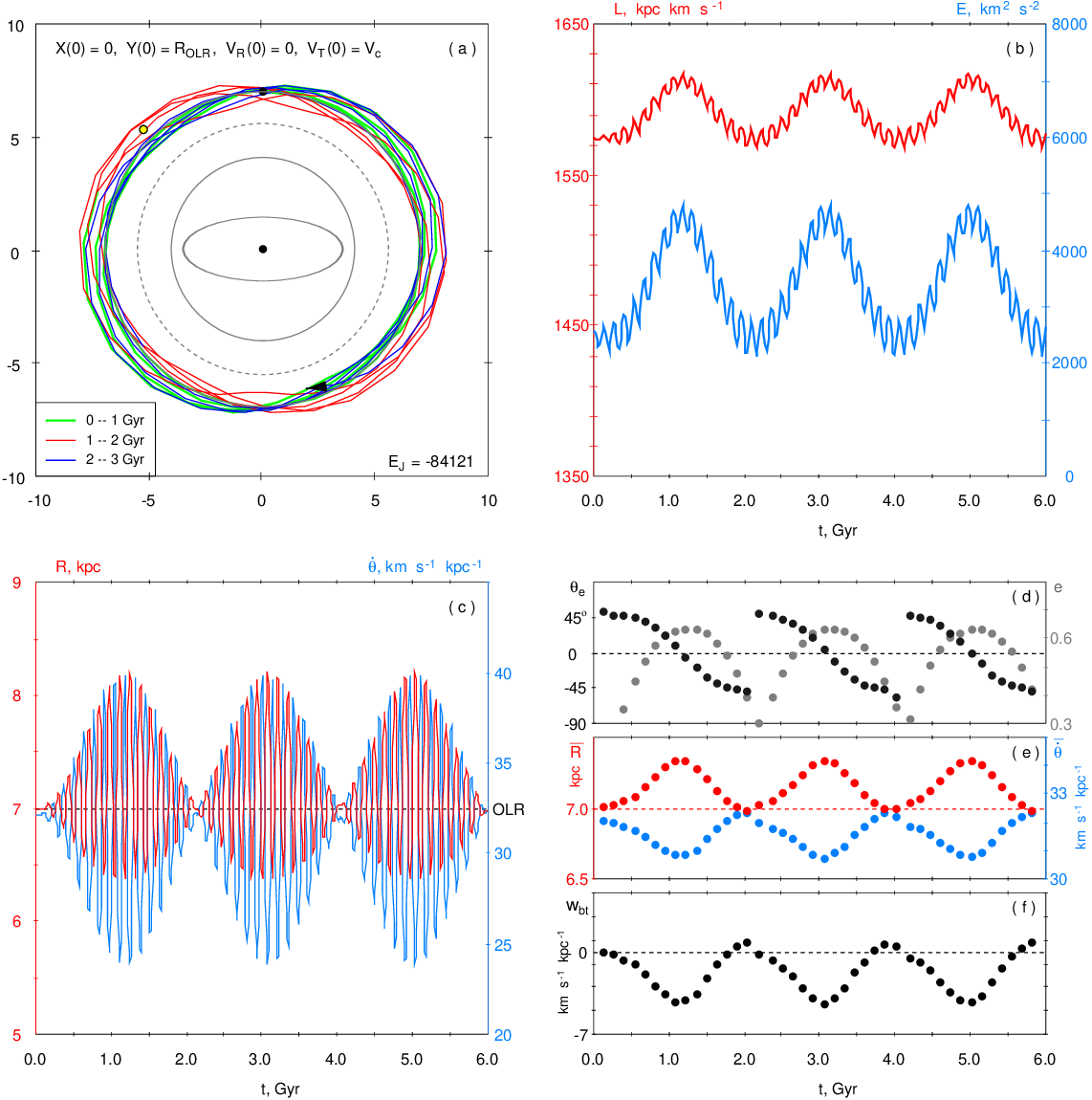}} \caption{(a) Orbit
of the star with the initial coordinates $X(0)=0$ and $Y(0)=R_{OLR}$
and initial velocities $V_R(0)=0$ and $V_T(0)=V_c$ in the reference
frame of the rotating bar.  Also shown is the Jacobi integral, $E_J$,
related to the time interval  $t>T_g$. The black and yellow circles
indicate the initial position of the star and the supposed position
of the Sun, respectively. The segments of the orbit for the time
periods 0--1, 1--2, and 2--3 Gyr are shown in green, red, and
dark-blue, respectively. The Galaxy rotates counterclockwise. The
arrow indicates the sense of  rotation of the star in the bar
reference frame. The position of the bar is shown by an ellipse. The
gray solid lines indicate the CR and OLR, and the dashed line, the
resonance $-4/1$. We can  see that the direction of orbit elongation
is slightly inclined to the right relative to the minor axis of the
bar during the time period 0--1 (green line), but it is inclined to
the left during the period 1--2 Gyr (red line), and the orientation
of the orbit returns to the initial position (tilted to the right)
during the period 2--3 Gyr (dark-blue line). (b) Variations in the
angular momentum $L$ of the star (red curve) and in the total energy
$E$ (blue curve) as a function of time. The left and right vertical
axes show the scales of changes in $L$ and $E$, respectively.  We can
clearly see short- and long-term oscillations in $L$ and $E$.
Long-term oscillations have the period of $P=1.91\pm0.01$ Gyr. (c)
Variations in the Galactocentric distance of the star $R$ (red curve)
and in its instantaneous angular velocity $\dot{\theta}$ (blue curve)
as functions of time. The scales of changes in  $R$ and
$\dot{\theta}$ are shown on the left and right axes, respectively. It
can be clearly seen that  oscillations in $R$ and $\dot{\theta}$
occur in anti-phase. The horizontal dashed  line indicates the
position of the OLR. (d) Variations in the direction of the orbit
elongation ($\theta_e$, black dots, left vertical axis), and in the
eccentricity of the orbit ($e$, gray dots, right vertical axis). The
angle $\theta_e$ is measured from the major axis of the bar and
increases in the sense of Galactic rotation. (e) Variations in the
average values of $\overline{R}$ and $\overline{\dot{\theta}}$. (f)
Variations in the beat frequency $w_{bt}$. } \label{orb_out_1}
\end{figure*}

In section 4.2 we  showed that the periodic enhancement of either
trailing or leading segments of the outer rings (oscillations of the
parameter $p_{max}$) occurs with the period of $2.0\pm0.1$ Gyr. Let
us consider an orbit which may be important for understanding the
cause of these changes. Fig.~\ref{orb_out_1} (a) shows the orbit of a
star with the initial coordinates $X(0)=0$ and $Y(0)=R_{OLR}$ and
initial velocities $V_R(0)=0$ and $V_T(0)=V_c$, where $V_c$ is the
velocity of the rotation curve at the radius of the OLR.  The orbit
is considered in the reference frame of the rotating bar. We can see
that in the time period 0--1 Gyr (green line) the direction of the
orbit elongation  is slightly inclined to the right relative to the
minor axis of the bar, but  in the time period 1--2 Gyr (red line) it
is inclined to the left, and in the period 2--3 Gyr (dark blue line)
the orbit returns to the original position (tilted to the right).
When the orbit is inclined to the right (0--1, 2--3 Gyr), it supports
the trailing segments of the rings $R_1$ and $R_2$, but when it is
inclined to the left (1--2 Gyr), it supports the leading segments of
the outer rings.

Note that the orbits considered in this section were obtained for the
bar that turns on slowly. However, this assumption is of no
particular importance in the case of the outer rings: both bar onset
scenarios result in nearly the same pattern of changes in the
orientation of orbits.

Fig.~\ref{orb_out_1} (a) also shows the supposed position of the Sun
which lags behind  the direction of the bar major axis by the angle
of $\theta_b=45^\circ$ (for more details, see Melnik et al. 2021).
Note that the orbits inclined to the right (green and dark-blue
lines) can cause  the appearance  of a large number of stars with
large radial velocities directed away from the Galactic center in the
solar vicinity which can appear and disappear in  certain time
periods.

Fig.~\ref{orb_out_1} (b) shows the variations in the specific angular
momentum $L$ (red curve) of the star and in the specific total energy
$E=K+\Phi$ (blue curve) as a function of time, where $K$ and $\Phi$
are the specific kinetic and potential energy of the star,
respectively. Since our model includes the isothermal halo, whose
potential grows indefinitely at infinity, we introduced a constant
into the formula  for $\Phi$, for  the potential energy at the
distance $R_{max}$ to be  $\Phi=0$. We integrate orbits up to the
distance of 11 kpc and therefore we adopted $R_{max}=11$ kpc.  We can
clearly see the short-term and long-term oscillations of $L$ and $E$.
The short-term oscillations are due to the fact that a star twice in
a period of a revolution around a bar gains and loses the angular
momentum and energy from the bar. The periods of the short- and
long-term oscillations are $P=0.13\pm0.01$ and $1.91\pm0.01$ Gyr,
respectively.

The presence of a bar  rotating with the angular velocity $\Omega_b$
in a galaxy leads to the fact that neither the angular momentum of a
star $L$ nor its energy $E$ are conserved, the only quantity that is
conserved is their combination in the form of the Jacobi integral:

\begin{equation}
E_J= E-\Omega_b L
 \label{jacobi}
\end{equation}

\noindent \citep[for example,][]{binney2008}. In our model, the
Jacobi integral $E_J$ at the distance of the OLR is conserved with a
relative accuracy of 10$^{-5}$.

Fig.~\ref{orb_out_1} (b) shows that the star gradually,  step by
step, acquires angular momentum and energy from the bar and then also
gradually loses them. Though our models are computed up to 3 Gyr, we
integrated the orbits up to 6 Gyr to show that changes in $L$, $E$,
and other parameters are periodic.

Fig.~\ref{orb_out_1} (c) shows the variations in the Galactocentric
distance $R$ of the star (red curve) and in its angular rotation
velocity $\dot{\theta}$ (blue curve). We can see that $R$ and
$\dot{\theta}$ oscillate in anti-phase. Note that as the amplitude of
oscillations increases, the average distance $R$ also increases. In
general, the pattern of oscillations resembles beats which arise in
the case of a superposition of two oscillations with close
frequencies. Near the resonance, we actually have two independent
oscillation sources with close frequencies: the frequency with which
a star meets the perturbation from the bar,
$2(\overline{\dot{\theta}}-\Omega_b)$, and the epicyclic frequency,
$\kappa$. Formally, the beat frequency is defined by the following
expression:

\begin{equation}
w_{bt}= \kappa(\overline{R})+2(\overline{\dot{\theta}}-\Omega_b),
 \label{w_bt}
\end{equation}

\noindent where the coefficient 1/2 is absent because of
consideration the fact that the amplitude of the beats reaches
extreme values twice per period. The  'plus' sign before $\kappa$ is
due to the fact that epicyclic motions occur in the sense opposite
that of galactic rotation. The problem is that both $\dot{\theta}$
and $\kappa$ depend on the distance $R$. Note that the beat
frequency, $w_{bt}$  (Eq.~\ref{w_bt}), is a particular case of the
frequency of the resonance, $\Omega_s$ (Eq.~\ref{omega_s}).

Is the epicyclic  approximation valid for this orbit?
Fig.~\ref{orb_out_1} (c) shows that the maximum amplitude of the
radial oscillations of the star is 0.9 kpc. This value determines the
maximum size of the epicycle in radial direction. Given that the
average radius of the orbit is $R=7.2$ kpc,  the size of the epicycle
is about an order  of magnitude smaller than the average radius of
the orbit. It means that the  epicyclic approximation is appropriate
in this case.

Fig.~\ref{orb_out_1} (e) shows variations in the average  distance
$\overline{R}$ and in the average angular velocity
$\overline{\dot{\theta}}$ as a function of time. Averaging was
performed over the intervals of one oscillation along the radius:
from one crossing of the radius $R=R_{OLR}$ with the negative radial
velocity, $V_R<0$, to another. Fig.~\ref{orb_out_1} (c) indicates
that the star crosses this radius in each radial oscillation. The
angular velocity $\dot{\theta}$ was also averaged at the same
intervals. Variations in $\overline{R}$ and $\overline{\dot{\theta}}$
occur  in the ranges of $\overline{R}=6.98$--7.33 kpc and
$\overline{\dot{\theta}}=30.73$--32.33 km s$^{-1}$ kpc$^{-1}$,
respectively. The period of  changes is $P=1.91\pm0.01$ Gyr.

Fig.~\ref{orb_out_1} (d) shows variations of the direction of orbit
elongation ($\theta_e$, black circles) and  eccentricity ($e$, gray
circles) calculated at the intervals of one radial oscillation. Note
that  angle $\theta_e$ is a particular case of the slow angle
variable, $\theta_s$ (Eq.~\ref{theta_s}). It can be clearly seen that
the angle $\theta_e$ changes non-uniformly: it lingers near
$+45^\circ$, decreases fast at $\theta_e=0$, and again lingers near
$-45^\circ$. Then  a fast orbital rearrangement occurs during which
$\theta_e$ jumps from $-45^\circ$ to $45^\circ$. At this time the
orbit has minimum eccentricity and minimum average radius
$\overline{R}$.  The angle $\theta_e$ is measured from the major axis
of the bar in the sense of Galactic rotation. Note that $\theta_e=0$
(orbit stretched in the direction of the bar major axis) corresponds
to  maximum average distance $\overline{R}$. When the angle
$\theta_e$ is near $+45^\circ$, the orbit is inclined to the right
and supports the trailing segments of the outer rings, but when
$\theta_e$ is near $-45^\circ$, the orbit is inclined to the left and
supports the leading segments of the outer elliptical rings. The
changes in the angle $\theta_e$ and  the eccentricity of the orbit
have periods of $P=1.95\pm0.04$ and $P=1.93\pm0.03$ Gyr,
respectively. The eccentricity of the orbit varies  in the range of
0.12--0.63.

The question is why  $\overline{R}$ reaches its maximum and does not
increase further. This may be  due to the orientation of the orbit at
the time of maximum $\overline{R}$, when  it becomes elongated along
the major axis of the bar. At this orientation of the orbit, the bar
begins to  subtract effectively  the angular momentum and energy from
the star on the orbital segments located inside the OLR radius. This,
in turn, causes a decrease in the average distance $\overline{R}$ and
a change in the orientation of the orbit. Note that the star at each
radial oscillation is located both inside and outside the OLR radius
which is seen well in  Fig.~\ref{orb_out_1}c.

\citet{struck2015a} showed that the ratio of the epicyclic frequency
to the angular velocity of the rotation curve:

\begin{equation}
M=\kappa/\Omega
 \label{M}
\end{equation}

\noindent depends on the eccentricity $e$ of the orbit. For a flat
rotation curve the relation between $M_{cor}$ and $M_0=M(e=0)$ has
the following form:

\begin{equation}
\begin{array}{c}
 M_{cor}/M_0= 1.0013 - 0.00439x + 0.0520x^2 + 0.0169x^3\\
 +0.00180x^4,\\
\end{array}
 \label{M_cor}
\end{equation}

\noindent where

\begin{equation}
x=\log_{10} (1 - e).
 \label{x}
\end{equation}

\noindent The correction that takes into account eccentricity
slightly increases   $\kappa$. The maximum correction for the orbit
considered is obtained for the eccentricity $e=0.63$ and amounts to
+0.51 km s$^{-1}$ kpc$^{-1}$. It is  small compared to the changes in
$\kappa$, $\kappa=43.60$--45.89 km s$^{-1}$ kpc$^{-1}$, caused by
changes in the average distance $\overline{R}$.

Fig.~\ref{orb_out_1} (f) shows variations  in the angular velocity of
the beats, $w_{bt}$. We  used the average values of
$\overline{\dot{\theta}}$ and $\kappa(\overline{R})$ in our
calculation of $w_{bt}$ (Eq.~\ref{w_bt}). It can be seen that
$w_{bt}$ reverses the sign, but the values of $w_{bt}$ are negative
most of the time. Note that the frequency $w_{bt}$ changes in
anti-phase with the average distance $\overline{R}$. Considering the
changes in $w_{bt}$ as a time series, we obtain the period of
$P=1.91\pm0.02$ Gyr.

Let us consider another orbit which demonstrates a variety of beat
patterns in the OLR region. Fig.~\ref{orb_out_2} (see Appendix) shows
the orbit of a star with the initial coordinates $X(0)=0$ and
$Y(0)=R_{OLR}$ and initial velocities $V_R(0)=0$ and $V_T(0)=V_c+15$
km s$^{-1}$. It also shows variations in the angular momentum of the
star $L$, total energy $E$,  instantaneous values of the distance
$R$, angular velocity $\dot{\theta}$, eccentricity $e$, average
distance $\overline{R}$,  average angular velocity
$\overline{\dot{\theta}}$, and frequency of beats $w_{bt}$ as a
function of time. The averaging of $\overline{R}$ and
$\overline{\dot{\theta}}$ was performed over time intervals from one
crossing of the radius $R=R_{OLR}+0.5$ kpc with a negative radial
velocity to another.

Fig.~\ref{orb_out_2} (e) shows that the direction of the orbit
elongation $\theta_e$ shifts  from +90 to $-90^\circ$ at a nearly
constant angular velocity. The elliptical orbit has the order of
symmetry equal to $m=2$ and therefore we are dealing  with the
rotation of the direction of the orbit elongation or with the
precession of the apsidal line. The period of the apsidal precession
can be estimated from the long-term changes in the angular momentum
and energy of the star which appears to be $P=0.81\pm0.01$ Gyr. In
this case, the changes in the average  distance $\overline{R}$ are
smaller and lie in the range $\overline{R}=7.46$--7.56 kpc. Note that
most of the time the star is located only outside the OLR, but during
the time periods when $\overline{R}$ reaches  maximum values, the
star moves inside the OLR for short time intervals
(Fig.~\ref{orb_out_2} c). The eccentricity of the orbit changes only
a little, $\overline{e}=0.42$--0.52, maximum correction of $\kappa$
due to orbital eccentricity is  +0.32 km s$^{-1}$ kpc$^{-1}$ which is
comparable to the  range of changes in the epicycle frequency,
$\Delta\kappa=0.63$ km s$^{-1}$ kpc$^{-1}$, caused by changes in
$\overline{R}$. The estimates of the period derived from variations
in the average distance $\overline{R}$, eccentricity, and beat
frequency $w_{bt}$ are also equal to $P=0.81\pm0.01$ Gyr.

The common feature of librating and precessing orbits is that they
both lie  inside and outside the OLR radius. The difference between
librating and precessing orbits is in the restriction on  angle
$\theta_e$, whose variations are restricted for librating orbits and
unrestricted for precessing ones (Weinberg 1994). In addition, we
found that librating orbits are characterized by  the reversal of the
sign of the beat frequency, $w_{bt}$, during a period of oscillations
while in the case of precessing orbits, the sign of $w_{bt}$  remains
unchanged. If $w_{bt}>0$ then the line of apsides rotates in the
sense of galactic rotation and if  $w_{bt}<0$, in the opposite sense.
A consequence of this feature is that  librating orbits have longer
oscillation periods than  precessing orbits.

Among orbits that lie both inside and outside the OLR in our model,
the fractions of librating  orbits is only  19 per cent, but a
maximum in the distribution of  librating orbits over the period
corresponds to  $P=1.9$ Gyr, and the fraction of  librating orbits
with the period in the range 1.8--2.0 Gyr amounts to $\sim60$ per
cent. Thus, our model includes a sufficient number of orbits with the
oscillation period of $\sim1.9$ Gyr.

\begin{figure*}
\resizebox{\hsize}{!}{\includegraphics{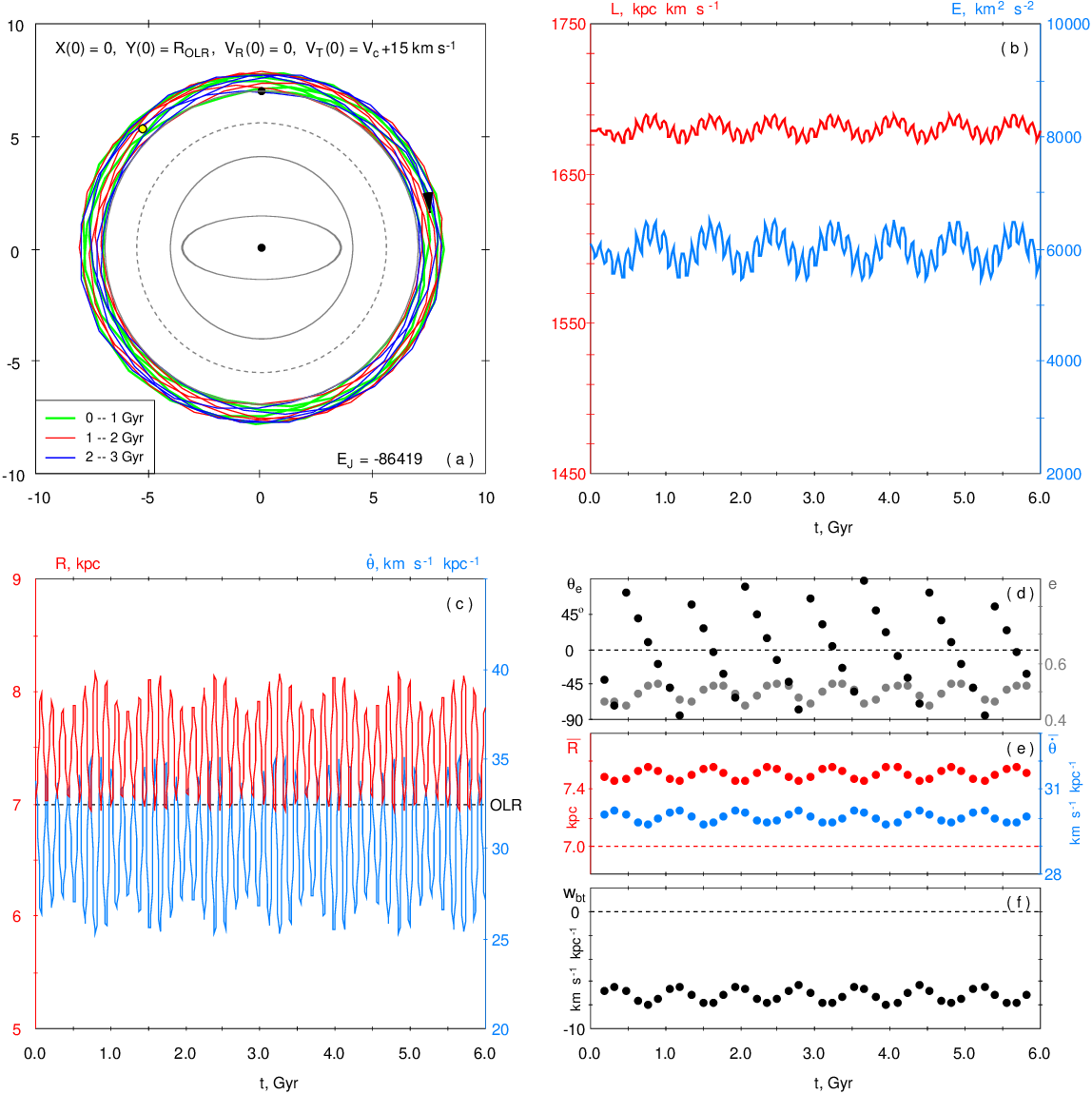}} \caption{(a) Orbit
of the star  with the initial coordinates $X(0)=0$ and $Y(0)=R_{OLR}$
and  velocities $V_R(0)=0$ and $V_T(0)=V_c+15$ km s$^{-1}$ in the
reference frame of the rotating bar.  Also shown is the Jacobi
integral, $E_J$. (b) Changes in the angular momentum of the star
($L$, red curve) and in the total energy ($E$, blue curve) as
functions of time. Long-term oscillations in $L$ and $E$ have a
period of $P=0.81\pm0.01$ Gyr. (c) Variations in the distance of the
star ($R$, red curve) and in its angular rotation velocity
($\dot{\theta}$, blue curve) as functions of time. The horizontal
dashed line indicates the position of the OLR. (d) Variations in the
direction of the orbit elongation, $\theta_e$, and eccentricity. It
can be observed  that  angle $\theta_e$ shifts nearly uniformly with
time. (e) Variations of average $\overline{R}$ and
$\overline{\dot{\theta}}$. (f) Variations of the beat frequency
$w_{bt}$. Note that $w_{bt}$ takes only negative values. For more
details, see caption to Fig.~\ref{orb_out_1}. } \label{orb_out_2}
\end{figure*}

\section{Discussion and conclusions}

We studied the periodic changes in the morphology of the resonance
rings using the dynamical model of the Galaxy which reproduces well
the distribution of observed velocities derived from the {\it Gaia}
EDR3 and {\it Gaia} DR3 catalogues along the Galactocentric distance.
The velocities calculated for  the {\it Gaia} EDR3 and {\it Gaia} DR3
data are consistent, on average, within 1 km s$^{-1}$.

We suppose that a sharp drop in the distribution of the azimuthal
velocity $V_T$  at $R\approx 4.0$--5.5 kpc (Fig.~\ref{obs_prof} c)
can be due to extinction, which  acts differently on thin-disc,
thick-disc, and halo stars. Note that the spiral distribution (snail
shell) of azimuthal velocities $V_T$ of {\it Gaia} DR2 and {\it Gaia}
DR3 stars in the ($Z$, $V_Z$) plane, which many authors attribute to
perturbations from the dwarf galaxy Sgr dSph \citep{antoja2018,
antoja2023}, can be due to similar effects.

The model disc forms the  nuclear ring, inner combined ring, and
outer resonance rings $R_1$ and $R_2$ (Fig.~\ref{distrib}).

We  demonstrated that  the periodic changes in the ring morphology
are not due to random deviations in the initial distributions of the
density and velocity (Fig.~\ref{amp_p}).

The inner ring is located in the region between the CR (4.0 kpc) and
the resonance $-4/1$ (5.5 kpc). This ring is not like other resonance
rings where stars make a complete revolution around the Galactic
center. The identified inner ring can be called a combined feature:
stars move along banana-type orbits near the stable equilibrium
points $L_4$ and $L_5$. We have found a density increase on  either
trailing or leading segments of the inner ring with a period of
$P=0.57\pm0.02$ Gyr (Fig.~\ref{amp_p} a, b). The period of a
revolution of stars in long-period orbits near the equilibrium point
$L_4$ ($L_5$) is $P=0.565\pm0.002$ Gyr,  which is very close to the
period of morphological changes in the inner ring.

A possible explanation of periodic enhancement of either leading or
trailing segments of the inner ring is the presence of the
overdensity, which forms at  the top point of banana-shaped orbits
when  the bar reaches its full strength and then  begins circulating
along the closed contour  (Fig.~\ref{overden}).

Orbits associated with the unstable equilibrium points L1 and L2 also
support the inner combined ring \citep[in agreement
with][]{athanassoula2009b}. Orbits associated with the unstable
equilibrium points $L_1$ and $L_2$ also support the inner combined
ring. In the case of slow bar onset a great variety of orbits
associated with the equilibrium points $L_1$ and $L_2$ transforms
into banana-shaped orbits (Fig.~\ref{orb_L1_all}).

How would we classify the inner combined ring, if we could have seen
our Galaxy from the outside? Perhaps we would classify it as an outer
lens $L$, which surrounds the bar like a rim. \citet[][Table 7
therein]{buta1996} found that the outer lenses are elongated
perpendicular to the bar in 74 per cent cases of galaxies considered.

The mechanisms causing   the morphological changes near the CR and
near the OLR are different. Near the CR we observe librations of
stars along banana-shaped orbits, but orbits themselves do not
librate, so we need overdensity  to explain morphological changes
here. However, near the OLR we observe  librations of orbits, so
there is no need for overdensity there.

In the region of the outer rings, we have found  periodic enhancement
of either leading or trailing segments of the outer rings with a
period of $P=2.0\pm0.1$ Gyr  (Fig.~\ref{amp_p} c, d). We studied
orbits in the vicinity of the OLR and  found oscillations of the
direction of orbit elongation, which seem to be due to the beats
between the frequency with which a star meets perturbations from the
bar, $2(\Omega-\Omega_b)$, and the epicyclic frequency $\kappa$.  The
beat frequency (Eq.~\ref{w_bt}) is a particular case of the frequency
of the resonance (Eq.~\ref{omega_s}).

Among many oscillating orbits, we have discovered those with the
oscillation period of  $P=1.91\pm0.01$ Gyr. If the orbit is inclined
in the sense  opposite that of Galactic rotation (to the right in
Fig.~\ref{orb_out_1}), it  supports the trailing segments of the
rings $R_1$ and $R_2$, but if it is tilted in the sense of Galactic
rotation (to the left in Fig.~\ref{orb_out_1}), it supports the
leading segments of the rings. It may be just these orbits ($P
\approx 1.9$ Gyr) that cause the enhancement of either trailing or
leading segments of the outer rings with the period of $P=2.0\pm0.1$
Gyr.

Oscillations in the orientation of the orbits are also accompanied by
long-term oscillations of the angular momentum $L$, energy $E$,
average  distance  $\overline{R}$, average angular velocity
$\overline{\dot{\theta}}$, eccentricity $e$, and beat frequency
$w_{bt}$ (Eq.~\ref{w_bt}). Oscillations  of these parameters have the
same periods. Maxima of the average distance $\overline{R}$
correspond to maxima of $L$, $E$, and eccentricity $e$, as well as to
minima of the average angular velocity $\overline{\dot{\theta}}$ and
beat frequency $w_{bt}$ (Fig.~\ref{orb_out_1}).

Among many factors that determine the oscillation period, it is worth
emphasizing the closeness of the average radius of the orbit  to the
radius of the OLR.

The angle $\theta_e$ determines the direction of orbit elongation
with respect to the major axis of the bar and increases in the sense
of Galactic rotation. The orbit with the oscillation period of
$P=1.91\pm0.01$ Gyr exhibits  non-uniform  variations of the angle
$\theta_e$: it  decreases  slowly  near  $+45^\circ$, decreases
rapidly near $\theta_e=0$, and again decreases slowly  near
$-45^\circ$, and then a fast rearrangement of the orbit occurs during
which $\theta_e$ jumps from $-45^\circ$ to $45^\circ$. Note that
during the orbital rearrangement, the average size of the orbit
$\overline{R}$ and its eccentricity have minimum values.

Note that the period of oscillations of  the maximum value,
$A_{max}$, of the Fourier amplitude $|A_2|$ is $P=0.97\pm0.02$ Gyr,
which is approximately half of the oscillation period of the
parameter $p_{max}=-m/\tan \gamma$ equal to $P=2.0\pm0.1$ Gyr
(Fig.~\ref{chi_per} b). Oscillations in the amplitude $A_{max}$ may
be due to  oscillations in the orientation of the orbit: when it is
elongated  at the angle $\theta_e \approx 45^\circ$ ($\theta_e
\approx -45^\circ$) it makes bridges between the trailing (leading)
segments of the rings $R_1$ and $R_2$, which causes trailing
(leading) spiral arms to become longer which, in its turn, must
increase the value of $A_{max}$. Angle $\theta_e$ takes values of
$\pm45^\circ$ twice per period and therefore the oscillation period
of $A_{max}$ must be nearly twice shorter than that  of $\theta_e$,
which is equal to $P=1.95\pm0.04$ Gyr.

Among orbits that lie both inside and outside the OLR in our model,
the fractions of librating and precessing orbits are  19 and 75 per
cent, respectively. Besides, 5 and 1 per cent of  orbits considered
oscillate within $\pm15^\circ$ with respect to the major and minor
axes of the bar, respectively.  Note that a maximum in the
distribution of librating orbits over the period corresponds to the
value of $P=1.9$ Gyr, and the fraction of  librating orbits with the
periods in the range 1.8--2.0 Gyr amounts to $\sim60$ per cent. Thus,
our model includes a sufficient number of orbits with  the
oscillation period of $\sim1.9$ Gyr.

A clearer understanding of the frequency distribution of the
oscillation periods requires an additional study that goes beyond the
scope of this work.

\section{acknowledgements}

{\small  We  thank the anonymous  editor and referee for fruitful
discussion especially in part concerning orbits in barred galaxies.
We thank Ralph Sch\"onrich for his remarks and suggestions concerning
orbits in the case of instant and slow bar onset (section 5.2). This
work has made use of data from the European Space Agency (ESA)
mission {\it Gaia} (\verb"https://www.cosmos.esa.int/gaia"),
processed by the {\it Gaia} Data Processing and Analysis Consortium
(DPAC, \verb"https://www.cosmos.esa.int/web/gaia/dpac/consortium").
Funding for the DPAC has been provided by national institutions, in
particular the institutions participating in the {\it Gaia}
Multilateral Agreement. E.N. Podzolkova is a scholarship holder of
the Foundation for the Advancement of Theoretical Physics and
Mathematics "BASIS" (Grant No. 21-2-2-44-1).}

\section{Data Availability}

{\small The   data   underlying   this   article   were derived from
sources in   the   public   domain:   VizieR at
\verb"https://vizier.u-strasbg.fr/viz-bin/VizieR"}

\end{document}